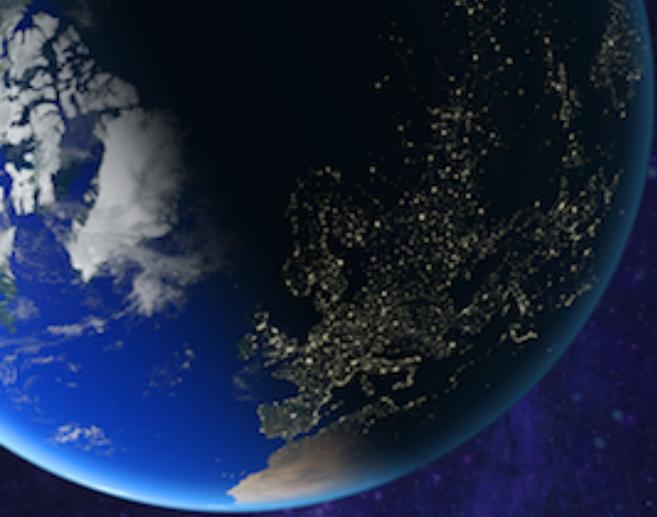

# LISA
# Laser Interferometer Space Antenna

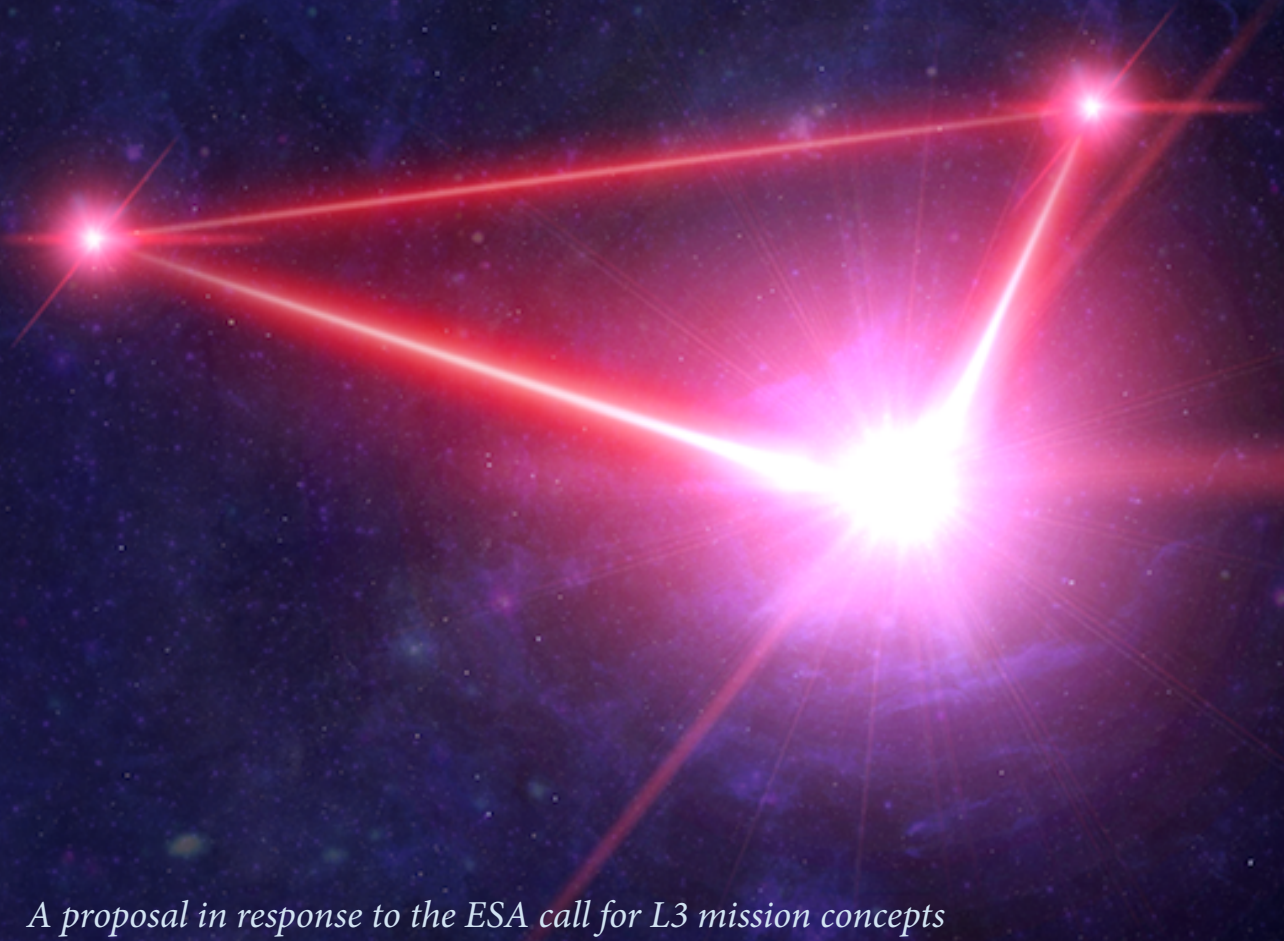

*A proposal in response to the ESA call for L3 mission concepts*

**Lead Proposer**
Prof. Dr. Karsten Danzmann

## Lead Proposer

Prof. Dr. Karsten Danzmann  karsten.danzmann@aei.mpg.de

Will be available with at least 20% of his time to support the study activities throughout the study.

Albert Einstein Institute Hannover
Leibniz Universität Hannover and Max Planck Institute for Gravitational Physics
Callinstr. 38, D-30167 Hannover, Germany

## Core Team

Pau Amaro-Seoane pau@ice.cat, Heather Audley heather.audley@aei.mpg.de, Stanislav Babak stba@aei.mpg.de, John Baker john.g.baker@nasa.gov, Enrico Barausse barausse@iap.fr, Peter Bender pbender@jila.colorado.edu, Emanuele Berti eberti@olemiss.edu, Pierre Binetruy binetruy@apc.univ-paris7.fr, Michael Born michael.born@aei.mpg.de, Daniele Bortoluzzi daniele.bortoluzzi@ing.unitn.it, Jordan Camp Jordan.B.Camp@nasa.gov, Chiara Caprini caprini@apc.in2p3.fr, Vitor Cardoso vitor.cardoso@ist.utl.pt, Monica Colpi Monica.Colpi@mib.infn.it, John Conklin jwconklin@ufl.edu, Neil Cornish cornish@physics.montana.edu, Curt Cutler curt.j.cutler@jpl.nasa.gov, Karsten Danzmann karsten.danzmann@aei.mpg.de, Rita Dolesi rita.dolesi@unitn.it, Luigi Ferraioli luigi.ferraioli@erdw.ethz.ch, Valerio Ferroni valerio.ferroni@unitn.it, Ewan Fitzsimons ewan.fitzsimons@stfc.ac.uk, Jonathan Gair j.gair@ed.ac.uk, Lluis Gesa Bote gesa@ice.csic.es, Domenico Giardini domenico.giardini@erdw.ethz.ch, Ferran Gibert ferran.gibert@unitn.it, Catia Grimani catia.grimani@uniurb.it, Hubert Halloin halloin@apc.univ-paris7.fr, Gerhard Heinzel gerhard.heinzel@aei.mpg.de, Thomas Hertog Thomas.Hertog@fys.kuleuven.be, Martin Hewitson martin.hewitson@aei.mpg.de, Kelly Holley-Bockelmann kelly.gravity@gmail.com, Daniel Hollington d.hollington07@imperial.ac.uk, Mauro Hueller mauro.hueller@unitn.it, Henri Inchauspe hinchaus@apc.in2p3.fr, Philippe Jetzer jetzer@physik.uzh.ch, Nikos Karnesis karnesis@aei.mpg.de, Christian Killow christian.killow@glasgow.ac.uk, Antoine Klein klein@iap.fr, Bill Klipstein William.Klipstein@jpl.nasa.gov, Natalia Korsakova natalia.korsakova@aei.mpg.de, Shane L Larson s.larson@northwestern.edu, Jeffrey Livas Jeffrey.Livas-1@nasa.gov, Ivan Lloro lloro@ice.csic.es, Nary Man nary.man@oca.eu, Davor Mance davor.mance@erdw.ethz.ch, Joseph Martino martino@apc.in2p3.fr, Ignacio Mateos mateos@ice.csic.es, Kirk McKenzie kirk.mckenzie@jpl.nasa.gov, Sean T McWilliams stmcwilliams@mail.wvu.edu, Cole Miller miller@astro.umd.edu, Guido Mueller mueller@phys.ufl.edu, Germano Nardini nardini@itp.unibe.ch, Gijs Nelemans nelemans@astro.ru.nl, Miquel Nofrarias nofrarias@ice.cat, Antoine Petiteau petiteau@apc.univ-paris7.fr, Paolo Pivato paolo.pivato@unitn.it, Eric Plagnol plagnol@apc.in2p3.fr, Ed Porter porter@apc.univ-paris7.fr, Jens Reiche jens.reiche@aei.mpg.de, David Robertson David.Robertson@glasgow.ac.uk, Norna Robertson nroberts@ligo.caltech.edu, Elena Rossi emr@strw.leidenuniv.nl, Giuliana Russano giuliana.russano@unitn.it, Bernard Schutz schutz@aei.mpg.de, Alberto Sesana asesana@star.sr.bham.ac.uk, David Shoemaker dhs@ligo.mit.edu, Jacob Slutsky jacob.p.slutsky@nasa.gov, Carlos F. Sopuerta sopuerta@ieec.uab.es, Tim Sumner t.sumner@imperial.ac.uk, Nicola Tamanini nicola.tamanini@cea.fr, Ira Thorpe james.i.thorpe@nasa.gov, Michael Troebs michael.troebs@aei.mpg.de, Michele Vallisneri michele.vallisneri@jpl.nasa.gov, Alberto Vecchio av@star.sr.bham.ac.uk, Daniele Vetrugno daniele.vetrugno@unitn.it, Stefano Vitale stefano.vitale@unitn.it, Marta Volonteri martav@iap.fr, Gudrun Wanner gudrun.wanner@aei.mpg.de, Harry Ward henry.ward@glasgow.ac.uk, Peter Wass p.wass@imperial.ac.uk, William Weber williamjoseph.weber@unitn.it, John Ziemer john.k.ziemer@jpl.nasa.gov, Peter Zweifel peter.zweifel@sed.ethz.ch

| Consortium Members | More than 300 scientists | https://www.lisamission.org/consortium/ |
| Supporters | More than 1300 researchers | https://www.lisamission.org/supporters/ |

Typeset: February 20, 2017

# Contents

# Executive Summary

The last century has seen enormous progress in our understanding of the Universe. We know that the Universe has emerged from the big bang, has been expanding at large, and contains luminous baryonic structures that shape our cosmic landscape. We know that stars are continuing to form in galaxies, and that galaxies form and assemble along filaments of the cosmic web. Powerful quasars and gamma-ray bursts were already in place when the Universe was less than one billion years old, indicating places where the first black holes formed. By using electromagnetic radiation as a tool for observing the Universe, we have learned that fluctuations at early epochs seeded the formation of all cosmic structures we see today. However, we do not know the nature of this dark component, which is revealed through its gravitational action on the luminous matter, nor how, when, and where the first black holes formed in dark matter halos.

We have come remarkably far using electromagnetic radiation as our tool for observing the Universe. However, gravity is the engine behind many of the processes in the Universe, and its action on all forms of mass and energy is dark. But gravity has its own messenger: Gravitational Waves, ripples in the fabric of spacetime, which travel essentially undisturbed from the moment of their creation. Observing Gravitational Waves from cosmic sources will let us explore a Universe inaccessible otherwise, a Universe where gravity takes on new and extreme manifestations.

The groundbreaking discovery of Gravitational Waves by ground-based laser interferometric Gravitational Wave observatories in 2015 is changing astronomy, giving us access to the high-frequency regime of Gravitational Wave astronomy. This is the realm of stellar mass objects at low redshift. Over the coming years, as the sensitivity of ground-based detectors improves, we will see the growth of a rich and productive Gravitational Wave astronomy. New sources with small mass will be discovered in the low redshift Universe. Already the first observation of Gravitational Waves brought a surprise, because the existence of such heavy stellar-origin binary black holes was not widely expected. But the low-frequency window below one Hertz will probably never be accessible from the ground. It is in this window that we expect to observe the heaviest and most diverse objects. Opening a gravitational window on the Universe in the low-frequency regime with the space-based detector LISA will let us go further than any alternative. These low-frequency waves let us peer deep into the formation of the first seed black holes, exploring redshifts larger than $z \sim 20$ prior to the epoch of cosmic re-ionisation, and examining systems of black holes with masses ranging from a few $M_\odot$ to $10^8\,M_\odot$. Exquisite and unprecedented measurements of black hole masses and spins will make it possible to trace the history of black holes across all stages of galaxy evolution, and at the same time test the General-Relativistic nature of black holes through detailed study of the amplitude and phase of the waveforms of Gravitational Wave strain. LISA will be the first ever mission to study the entire Universe with Gravitational Waves.

LISA is an all-sky monitor and will offer a wide view of a dynamic cosmos using Gravitational Waves as new and unique messengers to unveil *The Gravitational Universe*. It provides the closest ever view of the infant Universe at TeV energy scales, has known sources in the form of verification binaries in the Milky Way, and can probe the entire Universe, from its smallest scales near the horizons of black holes, all the way to cosmological scales. The LISA mission will scan the entire sky as it follows behind the Earth in its orbit, obtaining both polarisations of the Gravitational Waves simultaneously, and will measure source parameters with astrophysically relevant sensitivity in a band from below $10^{-4}$ Hz to above $10^{-1}$ Hz.

The LISA mission is proposed by an international collaboration of scientists called the LISA Consortium. Our proposal is fully compliant with the science goals indicated in the "Report of the Senior Survey Committee on the selection of the science themes for the L2 and L3 launch opportunities in the Cosmic Vision Programme". The team builds upon the proto-consortium that proposed a Gravitational Wave observatory for the L1 flight opportunity, and has been growing considerably ever since. It is augmented by additional member states and the US as an international partner. The LISA Consortium also proposed *The Gravitational Universe* as a science theme for the selection of the L2 and L3 launch opportunities and submitted the pertinent White Paper. The LISA Consortium also comprises all the investigators who have successfully pursued the LISA Pathfinder mission, a number of scientists who worked on the ground-based LIGO, Virgo, and GEO projects, and the Laser Ranging Interferometer on the GRACE Follow-On mission, thus making full use of all the expertise that has accumulated. This approach optimises the utilisation of the remaining time for mission preparation and technology development. We expect all mission elements to be at least at TRL 6 around 2020.

The LISA mission will be based on laser interferometry between free flying test masses inside drag-free spacecraft. These test masses, contained within the Gravitational Reference Sensors and effectively identical to the ones flown on LISA Pathfinder, will follow their geodesic trajectories with sub femto-g/$\sqrt{Hz}$ spurious acceleration. Two test masses free-fall inside each spacecraft, with each one serving as a geodesic reference end mirror for a single arm of the interferometer. The spacecraft is forced to follow the two test masses along each of the two interferometry axes they define, based on local interferometric position readouts. The test masses are then electrostatically suspended to the spacecraft along the other degrees of freedom, controlled by a combination of interferometric and capacitive position readouts. This system was successfully tested in the LISA Pathfinder mission, and this provides the confident basis for the acceleration performance of the mission.

The observatory will be based on three arms with six active laser links, between three identical spacecraft in a triangular formation separated by 2.5 million km. Continuously operating heterodyne laser interferometers measure with pm/$\sqrt{Hz}$ sensitivity in both directions along each arm, using well-stabilized lasers at 1064 nm delivering 2 W of power to the optical system. Again, using technology proven in LISA Pathfinder, the Interferometry Measurement System is using optical benches in each spacecraft. They will be constructed from an ultra-low expansion glass-ceramic to minimize optical pathlength changes due to temperature fluctuations. 30 cm telescopes transmit and receive the laser light to and from the other spacecraft. Three independent interferometric combinations of the light travel time between the test masses are possible, allowing, in data processing on the ground, the synthesis of two virtual Michelson interferometers plus a third null-stream, or "Sagnac" configuration.

The Consortium will deliver to ESA the integrated science instrument at the heart of the payload, plus several spacecraft-mounted parts of the instrument. It is expected that the remaining parts of the payload, in particular lasers and telescopes, will be procured by ESA or provided by NASA. The recommended option for LISA is to use one of the Ariane 6 family of launch vehicles, with a dedicated Ariane 6.4 launch being the preferred option. With a launch capacity directly into an escape trajectory of 7,000 kg, the Ariane 6.4 is very well suited to the LISA launch requirements into the LISA reference orbit, which is a stable Earth-trailing heliocentric orbit about 50 million km from Earth, with a mean inter-spacecraft separation of 2.5 million km. This reference orbit is optimised to minimise the key variable parameters of arm breathing angle and range rate between the spacecraft, as both of these drive the complexity of the payload design, while at the same time ensuring that the distance to the constellation is sufficiently small for communication purposes.

The entire constellation is expected to produce about 35 kbit/s of data in the nominal science mode, leading to a daily total of 334 MB. We augment the bidirectional laser links between the spacecraft with data links by modulating data onto the pseudo-random code used for ranging. Ground communication can then take place with only one of the three spacecraft per pass and still serve the whole constellation. With this configuration, for a single pointing of one antenna, communications can be maintained with a single ground station for 3 days at a user data rate of > 108.5 kbps for 7.2 hours of contact time per day using X band. This allows the re-pointing of the spacecraft antenna to happen once every 9 days (by cycling through the constellation), while still enabling daily communications with LISA to minimise data latency.

We propose a nominal mission duration of 4 years in science mode. However, the mission should be designed with consumables and orbital stability to facilitate a total mission of up to 10 years.

By 2030 our understanding of the Universe will have been dramatically improved by new observations of cosmic sources through the detection of electromagnetic radiation and high-frequency Gravitational Waves. Adding a low-frequency Gravitational Wave observatory will complement our astrophysical knowledge by using our new sense to 'hear' with low-frequency Gravitational Waves, providing access to a part of the Universe that will forever remain invisible with light. LISA will be the first ever mission to survey the entire Universe with Gravitational Waves. It will allow us to investigate the formation of binary systems in the Milky Way, detect the guaranteed signals from the verification binaries, alert astronomers of the imminent merger of heavy stellar-origin black holes, study the history of the Universe out to redshifts beyond $z = 20$, test gravity in the dynamical strong-field regime with unprecedented precision, and probe the early Universe at TeV energy scales. LISA will play a unique role in the scientific landscape of the 2030s.

# 1 Introduction

The groundbreaking discovery of Gravitational Waves (GWs) by ground-based laser interferometric detectors in 2015 is changing astronomy [1] by opening the high-frequency gravitational wave window to observe low mass sources at low redshift. The Senior Survey Committee (SSC) [2] selected the L3 science theme, *The Gravitational Universe* [3], to open the 0.1 to 100 mHz Gravitational Wave window to the Universe. This low-frequency window is rich in a variety of sources that will let us survey the Universe in a new and unique way, yielding new insights in a broad range of themes in astrophysics and cosmology and enabling us in particular to shed light on two key questions: (1) How, when and where do the first massive black holes form, grow and assemble, and what is the connection with galaxy formation? (2) What is the nature of gravity near the horizons of black holes and on cosmological scales?

We propose the LISA mission in order to respond to this science theme in the broadest way possible within the constrained budget and given schedule. LISA enables the detection of GWs from massive black hole coalescences within a vast cosmic volume encompassing all ages, from cosmic dawn to the present, across the epochs of the earliest quasars and of the rise of galaxy structure. The merger-ringdown signal of these loud sources enables tests of Einstein's General Theory of Relativity (GR) in the dynamical sector and strong-field regime with unprecedented precision. LISA will map the structure of spacetime around the massive black holes that populate the centres of galaxies using stellar compact objects as test particle-like probes. The same signals will also allow us to probe the population of these massive black holes as well as any compact objects in their vicinity. A stochastic GW background or exotic sources may probe new physics in the early Universe. Added to this list of sources are the newly discovered LIGO/Virgo heavy stellar-origin black hole mergers, which will emit GWs in the LISA band from several years up to a week prior to their merger, enabling coordinated observations with ground-based interferometers and electromagnetic telescopes. The vast majority of signals will come from compact galactic binary systems, which allow us to map their distribution in the Milky Way and illuminate stellar and binary evolution.

LISA builds on the success of LISA Pathfinder (LPF) [4], twenty years of technology development, and the Gravitational Observatory Advisory Team (GOAT) recommendations. LISA will use three arms and three identical spacecraft (S/C) in a triangular formation in a heliocentric orbit trailing the Earth by about 20°. The expected sensitivity and some potential signals are shown in Figure 1.

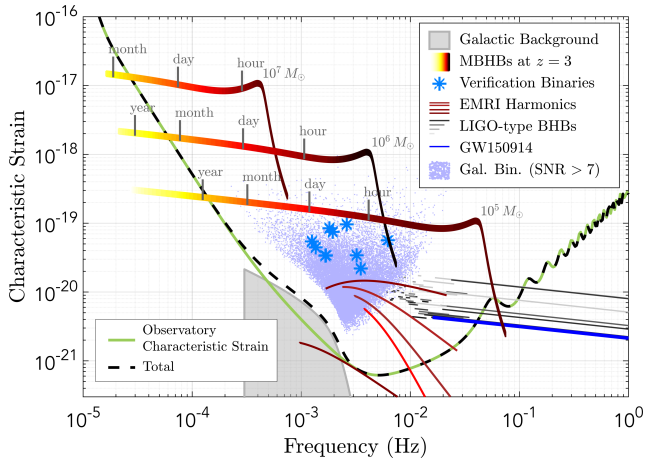

**Figure 1: Examples of GW sources in the frequency range of LISA, compared with its sensitivity for a 3-arm configuration.** The data are plotted in terms of dimensionless 'characteristic strain amplitude' [5]. The tracks of three equal mass black hole binaries, located at $z = 3$ with total intrinsic masses $10^7$, $10^6$ and $10^5 \, M_\odot$, are shown. The source frequency (and SNR) increases with time, and the remaining time before the plunge is indicated on the tracks. The 5 simultaneously evolving harmonics of an Extreme Mass Ratio Inspiral source at $z = 1.2$ are also shown, as are the tracks of a number of stellar origin black hole binaries of the type discovered by LIGO. Several thousand galactic binaries will be resolved after a year of observation. Some binary systems are already known, and will serve as verification signals. Millions of other binaries result in a 'confusion signal', with a detected amplitude that is modulated by the motion of the constellation over the year; the average level is represented as the grey shaded area.

An observatory that can deliver this science is described by a sensitivity curve which, below 3 mHz, will be limited by acceleration noise at the level demonstrated by LPF. Interferometry noise dominates above 3 mHz, with roughly equal allocations for photon shot noise and technical noise sources. Such a sensitivity can be achieved with a 2.5 million km arm-length constellation with 30 cm telescopes and 2 W laser systems. This is consistent with the GOAT recommendations and, based on technical readiness alone, a launch might be feasible around 2030. We propose a mission lifetime of 4 years extendable to 10 years for LISA.

## 2 Science performance

The science theme of *The Gravitational Universe* is addressed here in terms of Science Objectives (SOs) and Science Investigations (SIs), and the Observational Requirements (ORs) necessary to reach those objectives. The ORs are in turn related to Mission Requirements (MRs) for the noise performance, mission duration, *etc*. The majority of individual LISA sources will be binary systems covering a wide range of masses, mass ratios, and physical states. From here on, we use $M$ to refer to the total source frame mass of a particular system. The GW strain signal, $h(t)$, called the waveform, together with its frequency domain representation $\tilde{h}(f)$, encodes exquisite information about intrinsic parameters of the source (e.g., the mass and spin of the interacting bodies) and extrinsic parameters, such as inclination, luminosity distance and sky location. The assessment of Observational Requirements (ORs) requires a calculation of the Signal-to-Noise-Ratio (SNR) and the parameter measurement accuracy. The SNR is approximately the square root of the frequency integral of the ratio of the signal squared, $\tilde{h}(f)^2$, to the sky-averaged sensitivity of the observatory, expressed as power spectral density $S_h(f)$. Shown in Figure 2 is the square root of this quantity, the linear spectral density $\sqrt{S_h(f)}$, for a 2-arm configuration (TDI X). In the following, any quoted SNRs for the Observational Requirements (ORs) are given in terms of the full 3-arm configuration. The derived Mission Requirements (MRs) are expressed as linear spectral densities of the sensitivity for a 2-arm configuration (TDI X).

The sensitivity curve can be computed from the individual instrument noise contributions, with factors that account for the noise transfer functions and the sky and polarisation averaged response to GWs. Requirements for a minimum SNR level, above which a source is detectable, translate into specific MRs for the observatory. Throughout this section, parameter estimation is done using a Fisher Information Matrix approach, assuming a 4 year mission and 6 active links. For long-lived systems, the calculations are done assuming a very high duty-cycle ($> 95\%$). Requiring the capability to measure key parameters to some minimum accuracy sets MRs that are generally more stringent than those for just detection. Signals are computed according to GR, redshifts using the cosmological model and parameters inferred from the Planck satellite results, and for each class of sources, synthetic models driven by current astrophysical knowledge are used in order to describe their demography. Foregrounds from astrophysical sources, and backgrounds of cosmological origin are also considered.

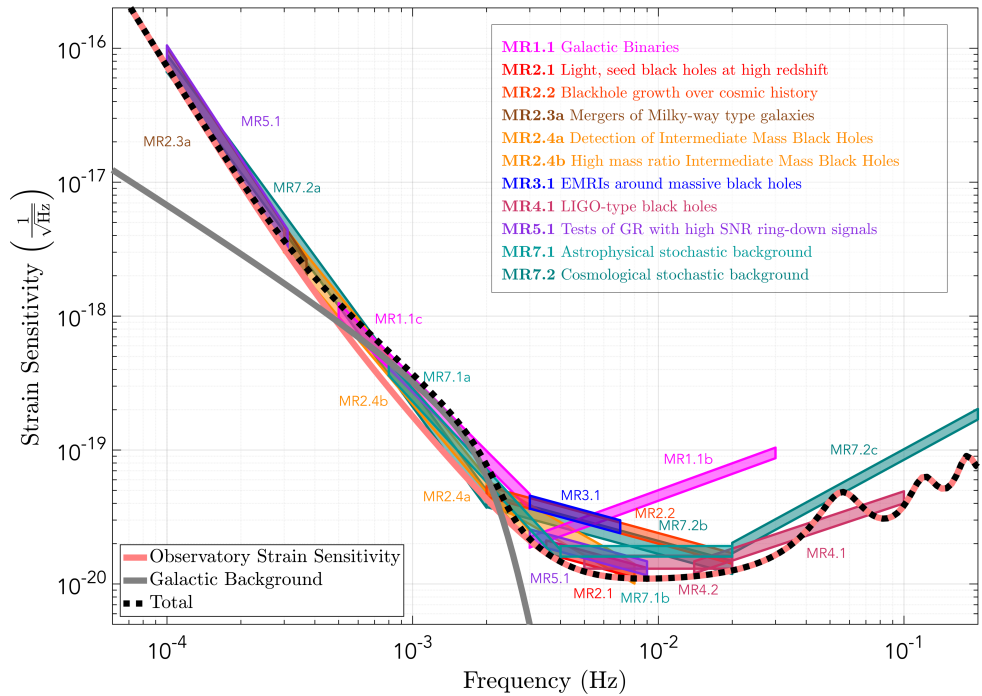

**Figure 2: Mission constraints on the sky-averaged strain sensitivity of the observatory for a 2-arm configuration (TDI X),** $\sqrt{S_h(f)}$, derived from the threshold systems of each observational requirement.

## 2.1 SO1: Study the formation and evolution of compact binary stars in the Milky Way Galaxy.

Numerous compact binaries in the Milky Way galaxy emit continuous and nearly monochromatic GW signals in the source frame [6]. These Galactic Binaries (GBs) comprise primarily white dwarfs but also neutron stars and stellar-origin black holes in various combinations. For those systems that can be detected, the orbital periods $P = 2/f$ can often be measured to high accuracy. The orbital motion of the detector imparts a characteristic frequency and amplitude modulation that allows us to constrain the extrinsic properties of some of the systems. Higher frequency systems are typically louder and better characterized than low frequency systems. At low frequencies, GBs are thought to be so numerous that individual detections are limited by confusion with other binaries yielding a stochastic foreground or confusion signal. Several "verification" binaries are currently known for which joint gravitational and electromagnetic (EM) observations can be done and many more will be discovered in the coming years, e.g., by Gaia and LSST. Using the current best estimate for the population [7], and assuming the reference sensitivity, it should be possible to detect and resolve ∼ 25,000 individual GBs.

**SI1.1: Elucidate the formation and evolution of GBs by measuring their period, spatial and mass distributions.**

*OR 1.1.a:* To survey the period distribution of GBs, and have the capability to distinguish between ∼ 5000 systems with inferred period precision $\delta P/P < 10^{-6}$.

*OR 1.1.b:* To measure the mass, distance and sky location for the majority of these GBs with frequency $f > 3$ mHz, chirp mass $> 0.2\,M_\odot$ and distance $< 15$ kpc.

*OR 1.1.c:* To detect the low frequency galactic confusion noise in the frequency band from 0.5 to 3 mHz. In Figure 2, the galactic confusion signal for a fiducial population is shown assuming a 4 year observation after subtraction of individual sources.

<u>MR1.1:</u> The ORs pose requirements in the band from about 0.5 mHz to 30 mHz. OR 1.1b demands that the sensitivity for frequencies 3 mHz $< f <$ 30 mHz has $\sqrt{S_h(f)} < 9 \times 10^{-21}(f/\text{mHz})^{2/3}$. For the frequency band indicated, this corresponds to having a strain sensitivity better than $1.2 \times 10^{-20}$ Hz$^{-1/2}$ at 3 mHz, and $7.8 \times 10^{-20}$ Hz$^{-1/2}$ at 30 mHz. From OR1.1.c, the identification of the low frequency galactic confusion signal requires us to be able to subtract all the identified/known sources with a certain precision which is limited by the other unknown sources as well as the detector sensitivity. In order for the detector sensitivity not to limit this significantly, we require the detector noise level below 2 mHz to be at, or below, the combined signal from galactic binaries. Using a conservative estimate for the galactic population sets a limit on the sensitivity in the band 0.5 mHz $< f <$ 3 mHz given by $\sqrt{S_h(f)} < 2.7 \times 10^{-19}(f/\text{mHz})^{-11/6}$. For the band discussed here, this corresponds to having a strain sensitivity better than $8.7 \times 10^{-19}$ Hz$^{-1/2}$ at 0.5 mHz, and $3.2 \times 10^{-20}$ Hz$^{-1/2}$ at 3 mHz.

**SI1.2: Enable joint gravitational and electromagnetic observations of GBs to study the interplay between gravitational radiation and tidal dissipation in interacting stellar systems.**

*OR 1.2.a:* To detect ∼ 10 of the currently known verification binaries, inferring periods with accuracy $\delta P/P < 10^{-6}$.

*OR 1.2.b:* To enable identification of possible electromagnetic counterparts, determine the sky location of ∼ 500 systems within one square degree.

*OR 1.2.c:* To study the interplay between gravitational damping, tidal heating, and to perform tests of GR, localise ∼ 100 systems within one square degree and determine their first period derivative to a fractional accuracy of 10% or better.

<u>MR1.2:</u> OR's 1.1.a, 1.1.b and 1.2.b,c set requirements on the mission duration in order to achieve the desired measurement precision. These requirements may not be fully met for mission durations less than 4 years.

## 2.2 SO2: Trace the origin, growth and merger history of massive black holes across cosmic ages

The origin of Massive Black Holes (MBHs) powering active nuclei and lurking at the centres of today's galaxies is unknown. Current studies predict masses for their *seeds* in the interval between about $10^3\,M_\odot$, and a few $10^5\,M_\odot$ and formation redshifts $10 \lesssim z \lesssim 15$ [8]. They then grow up to $10^8\,M_\odot$ and more by accretion episodes, and by repeated merging, thus participating in the clustering of cosmic structures [9], inevitably crossing the entire LISA frequency spectrum, from a few $10^{-5}$ Hz to $10^{-1}$ Hz, since their formation redshift. Mergers and accretion influence their spins in different ways thus informing us about their way of growing. The GW signal is transient, lasting from months to days down to hours. The signal encodes information on the inspiral and merger of the two spinning MBHs and the ring-down of the new MBH that formed. Being sources at cosmological redshifts, masses in the observer frame are $(1+z)$ heavier than in the source frame, and source redshifts are inferred from the luminosity distance $D_l$, extracted from the signal (with the exception of those sources for which we have an independent measure of $z$ from an identified electromagnetic counterpart). Consistent with current, conservative population models [10], the expected minimum observation rate of a few MBH Binaries (MBHB) per year would fulfill the requirements of SO2.

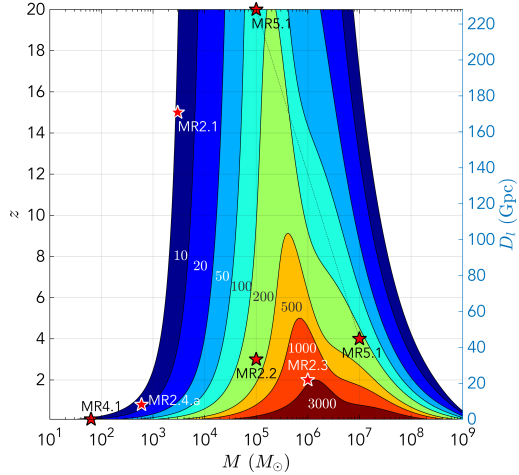

**Figure 3: Massive black hole binary coalescences:** contours of constant SNR for the baseline observatory in the plane of total source-frame mass, $M$, and redshift, $z$ (left margin-assuming Planck cosmology), and luminosity distance, $D_l$ (right margin), for binaries with constant mass ratio of $q = 0.2$. Overlaid are the positions of the threshold binaries used to define the mission requirements.

Figure 3 presents the richness of sources that should be visible to LISA, showing a wide range of masses observable with high SNR out to high redshift. The definition of the threshold systems (which are shown as red stars in Figure 3) for each OR leads to one or more MR, shown in Figure 2.

**SI2.1: Search for seed black holes at cosmic dawn**

*OR2.1* Have the capability to detect the inspiral of MBHBs in the interval between a few $10^3 \, M_\odot$ and a few $10^5 \, M_\odot$ in the source frame, and formation redshifts between 10 and 15. Enable the measurement of the source frame masses and the luminosity distance with a fractional error of 20% to distinguish formation models.

<u>MR2.1:</u> Ensure the strain sensitivity is better than $1.6 \times 10^{-20}$ Hz$^{-1/2}$ at 3.5 mHz and $1 \times 10^{-20}$ Hz$^{-1/2}$ at 9 mHz, to enable the observation of binaries at the low end of this parameter space with a SNR of at least 10. Such a "threshold" system would have a mass of 3000 $M_\odot$, mass ratio $q = 0.2$, and be located at a redshift of 15. All other MBHBs in OR2.1 with masses in the quoted range and mass ratios higher than this and/or at lower redshift, will then be detected with higher SNR yielding better parameter estimation.

**SI2.2: Study the growth mechanism of MBHs from the epoch of the earliest quasars**

*OR2.2.a* Have the capability to detect the signal for coalescing MBHs with mass $10^4 < M < 10^6 \, M_\odot$ in the source frame at $z \lesssim 9$. Enable the measurement of the source frame masses at the level limited by weak lensing (5 %).

*OR2.2.b* For sources at $z < 3$ and $10^5 < M < 10^6 \, M_\odot$, enable the measurement of the dimensionless spin of the largest MBH with an absolute error better than 0.1 and the detection of the misalignment of spins with the orbital angular momentum better than 10 degrees. This parameter accuracy corresponds to an accumulated SNR (up to the merger) of at least ~ 200.

<u>MR2.2:</u> The most stringent requirement is set by being able to measure the spin of a threshold system with total intrinsic mass of $10^5 \, M_\odot$, mass ratio of $q = 0.2$, located at $z = 3$. This will satisfy both OR2.1.a and 2.1.b. Achieving an SNR of 200 requires a strain sensitivity of $4 \times 10^{-20}$ Hz$^{-1/2}$ at 2 mHz and $1.3 \times 10^{-20}$ Hz$^{-1/2}$ at 20 mHz. All systems in OR2.2.a and 2.2.b with higher mass, mass ratios, spins, or lower redshift will result in higher SNR, and better spin estimation.

**SI2.3: Observation of EM counterparts to unveil the astrophysical environment around merging binaries**

*OR2.3.a* Observe the mergers of Milky-Way type MBHBs with total masses between $10^6$ and $10^7 \, M_\odot$ around the peak of star formation ($z \sim 2$), with sufficient SNR to allow the issuing of alerts to EM observatories with a sky-localisation of 100 deg$^2$ at least one day prior to merger. This would yield coincident EM/GW observations of the systems involved.

*OR2.3.b* After gravitationally observing the merger of systems discussed in OR2.3.a, the sky localisation will be significantly improved, allowing follow-up EM observations to take place. This has the potential to witness the formation of a quasar following a BH merger. This needs excellent sky localisation (about 1 deg$^2$) to distinguish from other variable EM sources in the field months to years after the merger.

<u>MR2.3:</u> For the lowest SNR system in OR2.3.a, which corresponds to a mass of $10^6 \, M_\odot$ at $z = 2$, we will detect the inspiral signal (with SNR=10) ~ 11.5 days prior to merger. Localising this source to 100 deg$^2$ requires an accumulated SNR of ~ 50, which will be known about 32 hours prior to merger if the strain sensitivity of the observatory is better than $7.2 \times 10^{-17}$ Hz$^{-1/2}$ at 0.1 mHz and $1.9 \times 10^{-18}$ Hz$^{-1/2}$ at 0.37 mHz. To achieve this operationally, data from the observatory need to be made available for analysis, around 1 day after measurement on-board. Additionally, in order to ensure coincident observations of GW and EM, we need to trigger a 'protected period' on-board during which no commissioning activities should take place. Hence there are three MRs here: a constraint on the strain sensitivity; a constraint on the cadence with which data are downloaded from the satellites; and the ability to trigger 'protected periods' where the instrument configuration is maintained. For all other systems in OR2.3.a with lower redshift, the SNR will be higher, and the sky-localisation correspondingly better.

**SI2.4 Test the existence of Intermediate Mass Black Hole Binaries (IMBHBs)**

*OR2.4.a:* Have the ability to detect the inspiral from nearly equal mass IMBHBs of total intrinsic mass between 600 and $10^4$ M$_\odot$ at $z < 1$, measuring the component masses to a precision of 30%, which requires a total accumulated SNR of at least 20.

<u>MR2.4.a:</u> Achieving a total SNR of about 20 for the systems described in OR2.4.a requires the strain sensitivity of the observatory to be better than $4.2 \times 10^{-20}$ Hz$^{-1/2}$ at 2 mHz and $1 \times 10^{-20}$ Hz$^{-1/2}$ at 8 mHz for the threshold system of 600 M$_\odot$ with a mass ratio of $q = 1$, located at $z = 1$.

*OR2.4.b:* Have the ability to detect unequal mass MBHBs of total intrinsic mass $10^4 - 10^6$ M$_\odot$ at $z < 3$ with the lightest black hole (the IMBH) in the intermediate mass range (between $10^2$ and $10^4$ M$_\odot$) [11], measuring the component masses to a precision of 10%, which requires a total accumulated SNR of at least 20.

<u>MR2.4.b:</u> Systems of OR2.4.b set constraints on the strain sensitivity of the observatory along the descending branch of the U-shaped curve where the galactic confusion noise-like signal dominates. Achieving a total SNR of 20 across that band for the systems described in OR2.4.b requires the strain sensitivity of the observatory to be better than $3 \times 10^{-18}$ Hz$^{-1/2}$ at 0.3 mHz, and $2 \times 10^{-20}$ Hz$^{-1/2}$ at 3 mHz. This requirement holds as long as the galactic confusion noise-like signal is at the level shown in Figure 2.

## 2.3 SO3: Probe the dynamics of dense nuclear clusters using EMRIs

Extreme Mass Ratio Inspirals (EMRIs) describe the long-lasting inspiral (from months to a few years) and plunge of Stellar Origin Black Holes (SOBHs), with mass range $10 - 60$ M$_\odot$, into MBHs of $10^5 - 10^6$ M$_\odot$ in the centre of galaxies [12]. The orbits of EMRIs are generic and highly relativistic. The SOBH spends $10^3 - 10^5$ orbits in close vicinity of the MBH, and the orbit displays extreme forms of periastron and orbital plane precession. The large number of orbital cycles allows ultra precise measurements of the parameters of the binary system as the GW signal encodes information about the spacetime of the central massive object. Considering the large uncertainty in the astrophysics of EMRIs, fulfillment of the requirements of this section would yield a minimum rate of one observed system per year, according to current most conservative EMRI population models.

**SI3.1 Study the immediate environment of Milky Way like MBHs at low redshift**

*OR3.1:* Have the ability to detect EMRIs around MBHs with masses of a few times $10^5$ M$_\odot$ out to redshift $z = 4$ (for maximally spinning MBHs, and EMRIs on prograde orbits) with the SNR $\geq 20$. This enables an estimate of the redshifted, observer frame masses with the accuracy $\delta M/M < 10^{-4}$ for the MBH and $\delta m/m < 10^{-3}$ for the SOBH. Estimate the spin of the MBH with an accuracy of 1 part in $10^3$, the eccentricity and inclination of the orbit to one part in $10^3$.

<u>MR3.1:</u> A threshold system for the range in OR3.1 would have a central non-spinning MBH with a mass of $5 \times 10^5$ M$_\odot$, a SOBH of 10 M$_\odot$ on a circular orbit, at redshift of 1.2. Such a system would have an accumulated SNR of 20 over a 4 year mission if the strain sensitivity of the observatory is better than $3.5 \times 10^{-20}$ Hz$^{-1/2}$ at 3 mHz and $2.3 \times 10^{-20}$ Hz$^{-1/2}$ at 7 mHz. All other systems with either lower redshift, higher component mass, or higher spin will produce a higher SNR. Systems with high spin and higher component mass may be detected out to redshift 4. Additionally we require the absence of any strong (SNR > 5) spectral lines of instrumental or environmental origin in the band from 2 to 20 mHz, which could interfere with the harmonics of the GW signal from these systems. The plunge time will be known to high accuracy several months ahead. It may also be useful to have the capability of triggering a protected period of about 1 week around the plunge time to allow testing the accumulation of SNR against GR.

## 2.4 SO4: Understand the astrophysics of stellar origin black holes

Following the LIGO discovery of SOBHs in the mass range from 10 to 30 $M_\odot$ merging in binary systems in the nearby Universe, a new science objective arises for LISA, which was not originally part of *The Gravitational Universe*. Based on the inferred rates from the LIGO detections, fulfillment of the requirements of this section would allow LISA to individually resolve a minimum number of about 100 SOBH binaries, some of which would cross into the LIGO band weeks to months later, enabling multi-band GW astronomy [13].

**SI4.1 Study the close environment of SOBHs by enabling multi-band and multi-messenger observations at the time of coalescence**

*OR4.1:* Have the ability to detect the inspiral signal from GW150914-like events with SNR > 7 after 4 years of observation and estimate the sky localisation with < 1 $\deg^2$ and the time of coalescence in ground-based detectors to within one minute. This will allow the triggering of alerts to ground-based detectors and to pre-point EM probes at the SOBH coalescence.

*MR4.1:* Detecting the inspiral of SOBHs with a mass comparable to those in the GW150914 system with SNR higher than 7, accumulated over 4 years, constrains the rising branch of the sensitivity curve by requiring a strain sensitivity of better than $1.2 \times 10^{-20}$ Hz$^{-1/2}$ at 14 mHz rising to $4 \times 10^{-20}$ Hz$^{-1/2}$ at 100 mHz.

**SI4.2 Disentangle SOBH binary formation channels**

*OR4.2:* Have the ability to observe SOBH binaries with total mass in excess of 50 $M_\odot$ out to redshift 0.1, with an SNR higher than 7 and a typical fractional error on the mass of 1 part in 100 and eccentricity with an absolute error of 1 part in $10^3$.

*MR4.2:* OR4.2 requires a strain sensitivity better than $1.3 \times 10^{-20}$ Hz$^{-1/2}$ between 5 and 20 mHz.

## 2.5 SO5: Explore the fundamental nature of gravity and black holes

MBHBs and EMRIs enable us to perform tests of GR in the strong field regime and dynamical sector [14, 15]. Precision tests such as these require 'Golden' binaries, that is, MBHBs with very high (> 100) SNR in the post-merger phase or EMRIS with SNR > 50.

**SI5.1 Use ring-down characteristics observed in MBHB coalescences to test whether the post-merger objects are the black holes predicted by GR.**

*OR5.1* Have the ability to detect the post-merger part of the GW signal from MBHBs with $M > 10^5$ $M_\odot$ out to high redshift, and observe more than one ring-down mode to test the "no-hair" theorem of GR.

*MR5.1:* The range of systems defined in OR5.1 sets a constraint on the sensitivity curve by requiring the high SNR and the observation of the merger. For masses at the low end of the range, the threshold system is one out at $z = 15$ with a mass of $10^5$ $M_\odot$, which will give an SNR of ~ 100 in the ringdown if the strain sensitivity is better than $2 \times 10^{-20}$ Hz$^{-1/2}$ at 3 mHz and $1 \times 10^{-20}$ Hz$^{-1/2}$ at 9 mHz. The contours of SNR in the mass/redshift plane are complicated, but we can constrain a point on the high mass end by considering a system of $10^7$ $M_\odot$ out at redshift 4. This system constrains the strain sensitivity to be better than $7 \times 10^{-17}$ Hz$^{-1/2}$ at 0.1 mHz, and $3 \times 10^{-18}$ Hz$^{-1/2}$ at 0.3 mHz, with the goal to extend this sensitivity down to low frequencies to see more of the inspiral phase, and allow earlier detection. Systems with masses between these two end points are considered 'Golden' binaries, yielding SNRs of up to 1000 for systems out to redshift 3.

**SI5.2 Use EMRIs to explore the multipolar structure of MBHs**

*OR5.2:* Have the ability to detect 'Golden' EMRIs (those are systems from OR3.1 with SNR > 50, spin > 0.9, and in a prograde orbit) and estimate the mass of the SOBH with an accuracy higher than 1 part in $10^4$, the mass of the central MBH with an accuracy of 1 part in $10^5$, the spin with an absolute error of $10^{-4}$, and the deviation from the Kerr quadrupole moment with an absolute error of better than $10^{-3}$.

*MR5.2:* The MRs are the same as MR3.1, but due to uncertainties in the astrophysical populations, a mission lifetime of several years is essential here to increase the chance of observing a Golden EMRI.

**SI5.3 Testing for the presence of beyond-GR emission channels**

Test the presence of beyond-GR emission channels (dipole radiation) to unprecedented accuracy by detecting GW150914-like binaries, which appear in both the LISA and LIGO frequency bands [16]. The ORs and MRs are the same as those in SI4.1.

**SI5.4 Test the propagation properties of GWs**

Test propagation properties of GW signals from EMRIs and from coalescing MBHBs. Detect the coalescence of Golden MBHBs (those systems described in OR2.2 with an SNR > 200) and have the ability to detect a Golden EMRI (as defined in OR5.2) which allows us to constrain the dispersion relation and set upper limits on the mass of the graviton and possible Lorentz invariance violations. The ORs and MRs are the same as those in MR2.2 and MR3.1.

**SI5.5 Test the presence of massive fields around massive black holes with masses $> 10^3 M_\odot$**

Constrain the masses of axion-like particles or other massive fields arising in Dark-Matter models by accurately measuring the masses and spins of MBHs [17]. The requirements on the accuracy of the mass and spin measurements are the same as in SI2.2.

Investigate possible deviations in the dynamics (encoded in the GW signal) of a solar mass object spiralling into an intermediate mass BH (mass < $a few 10^4 M_\odot$) due to the presence of a Dark Matter mini-spike around the IMBH [18]. This is a discovery project and the high frequency requirements stated in MR4.1, MR4.2 make such a discovery possible.

## 2.6 SO6: Probe the rate of expansion of the Universe

LISA will probe the expansion of the Universe using GW sirens at high redshifts: SOBH binaries ($z < 0.2$), EMRIs ($z < 1.5$), MBHBs ($z < 6$).

**SI6.1: Measure the dimensionless Hubble parameter by means of GW observations only**

*OR6.1a* Have the ability to observe SOBH binaries with total mass $M > 50\,M_\odot$ at $z < 0.1$ with SNR higher than 7 and typical sky location of $< 1\,\mathrm{deg}^2$.

*OR6.1b* Have the ability to localize EMRIs with an MBH mass of $5 \times 10^5\,M_\odot$ and an SOBH of $10\,M_\odot$ at $z = 1.5$ to better than $1\,\mathrm{deg}^2$.

<u>MR6.1:</u> In terms of sensitivity curve, the OR6.1a-b are automatically met if MR3.1 and MR4.1 are fulfilled. The need to collect a large enough sample of sources translates into a minimal mission duration requirement. According to current best population estimates, a 4 year mission is needed to yield a measurement of the Hubble parameter to better than 0.02, which helps resolving the tension among the values of the Hubble parameter determined with local Universe standard candles and with the Cosmic Microwave Background.

**SI6.2: Constrain cosmological parameters through joint GW and EM observations**

*OR6.2* Have the capability to observe mergers of MBHBs in the mass range from $10^5$ to $10^6\,M_\odot$ at $z < 5$, with accurate parameter estimation and sky error of $< 10\,\mathrm{deg}^2$ to trigger EM follow ups [19].

<u>MR6.2</u> In terms of the sensitivity curve, OR6.2 is automatically met if the MRs related to SO2 are fulfilled. The need to collect a large enough sample of sources translates into a minimal mission duration requirement. According to current best population estimates, a 4 year mission is needed to yield a measurement of the Hubble parameter to 0.01 and the dark energy equation of state parameter, $w_0$, to 0.1. A mission extension to 10 years would yield an improvement of a factor of about 2 on the measurement errors of these parameters.

## 2.7 SO7: Understand stochastic GW backgrounds and their implications for the early Universe and TeV-scale particle physics

One of the LISA goals is the direct detection of a stochastic GW background of cosmological origin (like for example the one produced by a first-order phase transition around the TeV scale) and stochastic foregrounds. Probing a stochastic GW background of cosmological origin provides information on new physics in the early Universe. The shape of the signal gives an indication of its origin, while an upper limit allows to constrain models of the early Universe and particle physics beyond the standard model.

For these investigations we need to ensure the availability of the data streams needed to form the Sagnac (or null-stream) TDI channel where the GW signal is partially suppressed in order to help separate the GW background from instrument noise.

**SI7.1: Characterise the astrophysical stochastic GW background**

*OR7.1:* Characterise the stochastic GW background from SOBH binaries with energy density normalised to the critical energy density in the Universe today, $\Omega$, based on the inferred rates from the LIGO detections, i.e., at the lowest $\Omega = 2 \times 10^{-10}\,(f/25\,\mathrm{Hz})^{2/3}$ [20]. This requires the ability to verify the spectral shape of this stochastic background, and to measure its amplitude in the frequency ranges $0.8\,\mathrm{mHz} < f < 4\,\mathrm{mHz}$ and $4\,\mathrm{mHz} < f < 20\,\mathrm{mHz}$.

<u>MR7.1:</u> The SNR over the 4 years of observation must be larger than 10 in the two frequency ranges. It would correspond to a strain sensitivity of better than $4 \times 10^{-20} (f/2.4\,\text{mHz})^{-2}\,\text{Hz}^{-1/2}$ for $0.8\,\text{mHz} < f < 4\,\text{mHz}$ (MR7.1a), and better than $1.6 \times 10^{-20}\,\text{Hz}^{-1/2}$ for $4\,\text{mHz} < f < 20\,\text{mHz}$ (MR7.1b).

### SI7.2 : Measure, or set upper limits on, the spectral shape of the cosmological stochastic GW background

*OR7.2:* Probe a broken power-law stochastic background from the early Universe as predicted, for example, by first order phase transitions [21] (other spectral shapes are expected, for example, for cosmic strings [22] and inflation [23]). Therefore, we need the ability to measure $\Omega = 1.3 \times 10^{-11} \left(f/10^{-4}\,\text{Hz}\right)^{-1}$ in the frequency ranges $0.1\,\text{mHz} < f < 2\,\text{mHz}$ and $2\,\text{mHz} < f < 20\,\text{mHz}$, and $\Omega = 4.5 \times 10^{-12} \left(f/10^{-2}\,\text{Hz}\right)^{3}$ in the frequency ranges $2\,\text{mHz} < f < 20\,\text{mHz}$ and $0.02 < f < 0.2\,\text{Hz}$.

<u>MR7.2:</u> Ensure an SNR higher than 10 over the 4 years of observation in the three frequency ranges specified in OR7.2 .

This would correspond to a strain sensitivity of better than $2.1 \times 10^{-19} (f/1\,\text{mHz})^{-2.5}\,\text{Hz}^{-1/2}$ (MR7.2a), $1.6 \times 10^{-20} (f/11\,\text{mHz})^{-0.5}\,\text{Hz}^{-1/2}$ (MR7.2b), and $9.3 \times 10^{-20} (f/0.11\,\text{Hz})\,\text{Hz}^{-1/2}$ (MR7.2c) in the ranges $0.1\,\text{mHz} < f < 2\,\text{mHz}$, $2\,\text{mHz} < f < 20\,\text{mHz}$ and $0.02\,\text{Hz} < f < 0.2\,\text{Hz}$, respectively.

*Additional remarks* Probing the gaussianity, the polarisation state, and/or the level of anisotropy of a potential stochastic background will give very important information about the origin of the background. In particular, limiting the number of instrumental glitches will help to assess the gaussianity. The polarisation state will be assessed with the 3 arm configuration. The measurement of the level of anisotropy depends on the frequency range and the amplitude of the background.

## 2.8 SO8: Search for GW bursts and unforeseen sources

LISA will lead us into uncharted territory, with the potential for many new discoveries. Distinguishing unforeseen, unmodelled signals from possible instrumental artifacts will be one of the main challenges of the mission, and will be crucial in exploring new astrophysical systems or unexpected cosmological sources.

### SI8.1: Search for cusps and kinks of cosmic strings

Searching for GW bursts from cusps and kinks of cosmic strings requires a deep understanding of the instrument noise and non-stationary behavior. Using the known shape of the bursts in the time and frequency domains will help to distinguish them from the instrumental artifacts and fluctuations in the stationarity of the instrument noise floor. Having the ability to use the Sagnac (or null-stream) TDI channels (MR7.2) to veto such instrumental events will play a crucial role in the exploration of this discovery space.

### SI8.2: Search for unmodelled sources

Searching for GW bursts from completely unmodelled and unforeseen sources will also require a deep understanding of the instrument noise and non-stationary behavior. To distinguish such signals from instrumental effects, it is essential that sources of instrumental non-stationary artifacts be kept as few as possible and that we maintain the ability to form the Sagnac combination (which is insensitive to GWs at low frequencies). This requires that we maintain 6 laser links for the full duration of the mission (MR7.2) and that we make available the necessary data streams to allow required computations on ground. This will help to veto out all non-GW burst-like disturbances.

## 2.9 Summary

LISA is a mission of discovery. Revealing *The Gravitational Universe* in LISA's frequency band will undoubtedly greatly enhance our knowledge of the Universe. Apart from the observation of the known "verification binaries", a lot of the science presented here depends on various models of astrophysical populations. Indeed, one of the primary goals of LISA is to constrain those population models. The baseline sensitivity proposed is one which is considered both technically feasible and at a level sufficient to achieve the science of *The Gravitational Universe*, given the current state of astrophysical population models.

The science addressed by LISA is extremely rich and covers many different domains of astrophysics. There is therefore not a single criterion for success of the mission, but success criteria for these various aspects of LISA Science. They are summarized in Table 1. The Mission Requirements (MRs) laid out above, which collectively specify a robust strain sensitivity level at frequencies between 0.1 mHz and 100 mHz over a science lifetime of at least 4 years with additional provisions for data latency and protected observing periods, define the sensitivity envelope required for complete mission success on all Science Investigations (SIs). If any of these MRs are not met, there is a graceful degradation of the science performance, that will affect differently the various SIs. Alternatively, if any of these MRs are exceeded, the mission may be able to outperform some of its Observational Requirements (Ors).

As Figure 1 shows, many of the LISA sources have extremely high signal-to-noise ratio (SNR) and are hardly affected by slight changes of the sensitivity curve. On the other hand, the threshold systems for the Observational Requirements (ORs) defining the envelope of the required sensitivity curve obviously are very sensitive to variations of the sensitivity curve.

The various signals used in the SIs to answer the different science questions are affected by different regions of the sensitivity curve. Also, some signals are affected more than others by the mission lifetime.

For example, the detectability of the SOBHs of SO4 is strongly affected by the high-frequency ($f \gtrsim 10\,\text{mHz}$) performance of the observatory: assuming the current estimate for the abundance of these sources, the baseline configuration should allow us to see on the order of 100 of these systems over the nominal 4 year mission. Improvements in high-frequency sensitivity of a factor of 2 would yield about a factor of 3 more detectable sources, whereas loss in sensitivity in this band by a factor of 2 will result in very few detections. Similarly, increasing the mission duration allows us to observe these systems for longer times, accumulate SNR, and enables us to detect more systems.

Another example, at the low-frequency end of the measurement band, arises from the ability to trigger electro-magnetic follow-up observations. MR2.3 spells this out for the baseline configuration, but the results are strongly dependent on the low-frequency performance. An improvement of the low-frequency performance by a factor of 4 pushes out the possible alert time from 1.5 weeks to 4 weeks, whereas a loss of low frequency performance by a factor of 4 will reduce the advance alert time to 2 days.

Looking at the most sensitive band of the observatory, a number of sources are affected by the level of sensitivity there. For example, the ability to make high precision measurements of parameters of MBHBs, such as the spin, requires very high SNRs of around 200. The number of systems we will be able to see with such SNRs will depend very strongly on the 'reach' of the observatory. Given the baseline, systems such as those in SI2.2 will be observable out to redshift 20 with sufficient SNR. Reducing the sensitivity of the observatory in the ~ 10 mHz frequency range reduces the observatory reach, and hence the number of potentially observable systems.

During the Phase 0 and Phase A studies, the trade-off between these non-independent requirements needs to be carefully examined to achieve the optimum science performance within a constrained budget. To support this trade-off, the consortium is preparing a science metric document that quantitatively identifies how some of the science is affected by graceful degradation or improved performance.

| *The Gravitational Universe* Objectives | Mission Success Criteria |
| --- | --- |
| Trace the formation, growth, and merger history of massive black holes | Perform SIs for **SO2** |
| Explore stellar populations and dynamics in galactic nuclei | Perform SIs for **SO3** |
| Test GR with observations | Perform SIs for **SO5** |
| Probe new physics and cosmology | Perform SIs for **SO6**, **SO7** and **SO8** |
| Survey compact stellar-mass binaries and study the structure of the Galaxy | Perform SIs for **SO1** and **SO4** |

**Table 1:** Association of mission success criteria with the science objectives as summarized in the *The Gravitational Universe* theme document.

# 3 Mission Profile

GWs change the light travel time or the optical pathlength between free falling [24] test masses (TMs). These test masses and the surrounding Gravitational Reference Sensor (GRS) hardware will exploit the full flight heritage of the same systems used on LISA Pathfinder. The test masses will follow their geodesic trajectories with sub-femto $g/\sqrt{\text{Hz}}$ spurious acceleration. They will be located inside three identical S/C in a triangular formation separated by 2.5 million km. Laser interferometers (IFOs) will measure the pm to nm pathlength variations caused by GWs. The interferometers are all-sky monitors of GWs and do not require nor allow for any pointing towards specific sources. The constellation will follow its initial orbit with very little to no orbital corrections and can observe continuously.

## 3.1 Orbit

The proposed orbit for LISA is an Earth-trailing heliocentric orbit between 50 and 65 million km from Earth,

with a mean inter-S/C separation distance of 2.5 million km. A reference orbit has been produced, optimised to minimise the key variable parameters of inter-S/C breathing angles (fluctuations of vertex angles) and the range rate of the S/C, as both of these drive the complexity of the payload design, while at the same time ensuring the range to the constellation is sufficiently close for communication purposes.

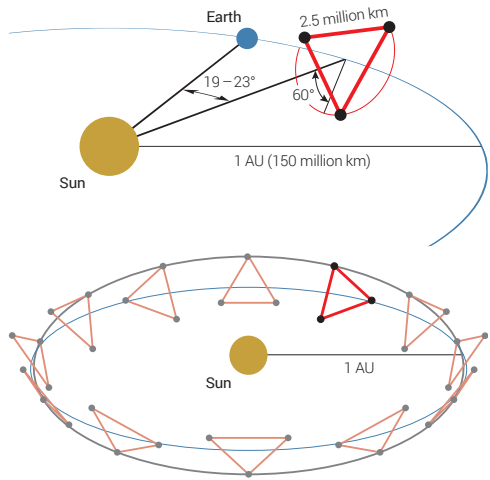

**Figure 4:** Depiction of the LISA Orbit.

The orbital configuration is depicted in Figure 4. These orbits will lead to breathing angles of ±1 deg and Doppler shifts between the S/C of within ±5 MHz.

The launch and transfer are optimized for a dedicated Ariane 6.4 launch, and carry the following basic features:

- total transfer time of about 400 days;
- direct escape launch with $V_\infty$ = 260 m/s;
- three sets of manoeuvres for final transfer orbit injection performed by the propulsion and S/C composite modules. See Section 5.4.3 for details.

## 3.2 Launcher

The recommended option for LISA is to use one of the Ariane 6 family of launch vehicles, with a dedicated Ariane 6.4 launch being the preferred option. With a launch capacity directly into an escape trajectory of 7,000 kg, the Ariane 6.4 is very well suited to the LISA launch requirements and the reference orbit described in Section 3.1 is based on the capabilities of this launcher. The capacity of Ariane 6.2 is limited, and it is extremely likely that any mission sized to fit within it would be significantly compromised in terms of capability. Similarly, it is likely that the constraints and complexity of a launch to Geostationary Transfer Orbit, combined with the need to find a suitable partner, make a shared Ariane 6.4 launch unattractive.

## 3.3 Concept of Operations

Each S/C is equipped with its own propulsion module to reach the desired orbit. During this cruise phase, checkout and testing of some equipment could already begin. Once the S/C have been inserted into their correct orbits and the propulsion modules jettisoned, the three S/C must be prepared to form a single working observatory before science operations can be established. This includes the release of the test masses and engaging the Drag-Free Attitude Control System (DFACS). This process, constellation acquisition and calibration, is described in Section 4.4.1. Following acquisition and calibration, LISA would enter the primary science mode. At this time, all test masses inside the three S/C will be in free fall along the lines of sight between the S/C. Capacitive sensors surrounding each test mass will monitor their position and orientation with respect to the S/C. DFACS will use micro-Newton thrusters to steer the S/C to follow the test masses along the three translational degrees-of-freedom, using interferometric readout where available, and capacitive sensing for the remaining degrees-of-freedom. Electrostatic actuators are used to apply the required forces and torques in all other degrees of freedom to the test masses. Laser interferometry is used to monitor the distance changes between the test masses and the optical bench (OB) inside each S/C. These technologies have been demonstrated by the LISA Pathfinder mission.

The long-baseline laser interferometer or science interferometer is used to measure changes in the distance between the optical benches while a third interferometer signal monitors the differential laser frequency noise between the two local laser systems. All interferometer signals are combined on ground to determine the differential distance changes between two pairs of widely separated test masses. Science Mode would feature near-continuous operation of the system at the design sensitivity. The system design should be such that, in science mode, external perturbations to the system are minimised and in particular the baseline design does not require station keeping or orbit correction manoeuvres. In line with the science requirements on data latency, communications would occur once per day for a duration of approximately 8 hours. There are two principal events which will cause some disruption to the science mode of operations; these are re-pointing of the antennas and re-configuration of the laser locking to maintain the beat notes within the phasemeter bandwidth, these are covered in more detail in Sections 3.5 and 4.4 respectively. In addition to the main science mode, a special protected period mode is envisaged. As identified in Section 2, there will be occasions when it is possible, with advance notice of around one week, to predict the time of a specific merger event. In this case, the protected period mode would be triggered to ensure that, over the ~ 1 day around the merger event there were no disruptions to the system. In particular, this would mean no antenna re-pointing and no laser frequency switching. This could be achieved by applying 1-2 days margin to the planned switching intervals, such that a planned re-pointing/switching could be moved out of the protected period.

### 3.4 Mission Lifetime

We propose a nominal mission duration of 4 years in science mode. Within this time, the key science requirements can be addressed to a suitable level as discussed in Section 2. Given the revolutionary and unique nature of LISA, however, the mission should be designed with consumables (e.g., DFACS propellant, available power) and orbital stability to facilitate a total mission up to 10 years in duration.

### 3.5 Communication requirements and strategy

The entire constellation is expected to produce about 35 kbit/s of data in the nominal science mode, as described in Section 5, Table 7, leading to a daily total of 334 MB. We augment the bidirectional laser links between each S/C with data links (around 15 kbit/s bidirectionally) by modulating data on the pseudo-random code used for ranging. This has been demonstrated with representative power levels at AEI and JPL [25, 26, 27]. Ground communication could then take place with only one of the three S/C per pass and still serve the whole constellation. With this configuration, it has been calculated that for a single pointing of one antenna, communications can be maintained with a single ground station for 3 days at a user data rate of > 108.5 kbps using X band, see Section 5.4.3. This allows the re-pointing of the antenna to happen once every 9 days (by cycling through the constellation), while still enabling daily communications with LISA to minimise data latency. At a rate of ≥ 108.5 kbps, and with a daily communications schedule, the complete 334 MB set of nominal data can be transmitted in < 7.2 hours. For a single station of the ESA ground network the minimum contact time per day to the LISA orbit has been calculated to be around 8 hours - sufficient for the nominal science data stream. Additionally, by utilising multiple ground stations (New Norcia, Cebreros and Malargue) the contact window could be extended to > 23 hours a day. While not the baseline, this option could be useful for calibration and commissioning operations.

## 4 Model Payload

### 4.1 Description of the measurement technique

LISA will detect gravitational waves with an interferometric measurement of differential optical pathlength modulation along the three sides of a triangular configuration defined by free-falling test masses, which are contained inside co-orbiting drag-free spacecraft. The distance changes between the test masses caused by the GWs are small (pm to nm) compared to the variations caused by solar system celestial dynamics (some 10000 km), but can be distinguished because the former are at mHz frequencies (1000 seconds timescale), whereas the latter have periods of many months and are quiet at mHz frequencies.

The optical pathlength measurement uses continuously operating heterodyne laser interferometers in both directions along each arm, using stable lasers at 1064 nm and a few Watts of power transmitted at each end. The beam divergence over several million km limits the received laser light power to some 100 pW, which rules out passive reflection for the return path. Instead, each S/C acts as an active transponder, transmitting a fresh high-power beam that is phase-locked to the incoming weak beam, with a fixed offset frequency. The constellation is fully symmetric, with similar measurements taking place in both directions along each of the three arms.

Three independent interferometric combinations of the light travel time measurements between the test masses are possible, allowing, in data processing on ground, the synthesis of two virtual Michelson interferometers plus a third ("Sagnac") configuration that is largely insensitive to GWs. The two independent Michelson interferometers allow simultaneous measurement of the two possible polarisations of the GW, and the Sagnac combination can be used to characterise the instrumental noise background. The yearly rotation of the constellation about itself and its orbit around the Sun allows to reconstruct the source direction on the sky for sources that can be observed for at least several weeks.

Noisy non-gravitational forces acting on the spacecraft require the use of test masses as geodesic reference test particles, which are shielded by the containing S/C. Two TM per spacecraft are used, each one dedicated to a single interferometry arm. To limit the relative S/C – TM accelerations, the spacecraft are "drag-free controlled" with micro-Newton thrusters to follow each TM along its interferometry arm, with no forces applied to the TM along these measurement axes. The total TM-TM measurement along each arm is separated into three parts:

- TM1 (test mass 1) to optical bench in S/C 1 (local);
- optical bench in S/C 1 to optical bench in S/C 2 through telescopes (long arm); and
- optical bench in S/C 2 to TM2 (local).

As orbital dynamics gives rise to relative velocities of order ±5 m/s between the S/C, the interferometric phase measurement system will have to track ~ 5 MHz-frequency Doppler shifts in the long, S/C to S/C interferometry measurement. Combining these three measurements in post-processing on ground will yield the desired TM to TM separation, and further postprocessing by the Time-Delay Interferometry [28] (TDI) algorithm will remove the otherwise dominating laser frequency noise by synthesizing virtual equal-armlength interferometers. The absolute inter-spacecraft distances are determined to the required ~ 10 cm accuracy using an auxiliary modulation on the laser beams.

## 4.2 Key measurement performance requirement

The strain sensitivity curve shown in Section 2 is determined by three main parameters listed below. We propose to set requirements above 0.1 mHz to limit the effort of testing but furthermore require that no features of the design shall preclude reaching the goal sensitivity down to 20 $\mu$Hz.

*Stray accelerations of the geodesic reference TM.* The proposed requirement is

$$S_a^{1/2} \leq 3 \cdot 10^{-15} \frac{\mathrm{m\,s^{-2}}}{\sqrt{\mathrm{Hz}}} \cdot \sqrt{1 + \left(\frac{0.4\,\mathrm{mHz}}{f}\right)^2} \cdot \sqrt{1 + \left(\frac{f}{8\,\mathrm{mHz}}\right)^4}$$

$$100\,\mu\mathrm{Hz} \leq f \leq 0.1\,\mathrm{Hz} \quad \mathrm{req.}$$
$$20\,\mu\mathrm{Hz} \leq f \leq 1\,\mathrm{Hz} \quad \mathrm{goal}$$

where $S_a^{1/2}$ is the single TM acceleration noise level.

Note that the conversion from acceleration to displacement gives the $1/f^2$ slope in the sensitivity graph (see Figure 2). This requirement mostly applies to the GRS that comprises the TM and the surrounding sensing and actuation hardware. The quoted level corresponds to what has been demonstrated by the LISA Pathfinder differential acceleration performance, with a little margin included. Further details and derived requirements are described below.

*Displacement noise of the interferometric TM-to-TM ranging* with a proposed requirement of

$$S_{\mathrm{IFO}}^{1/2} \leq 10 \cdot 10^{-12} \frac{\mathrm{m}}{\sqrt{\mathrm{Hz}}} \cdot \sqrt{1 + \left(\frac{2\,\mathrm{mHz}}{f}\right)^4}$$

$$100\,\mu\mathrm{Hz} \leq f \leq 0.1\,\mathrm{Hz} \quad \mathrm{req.}$$
$$20\,\mu\mathrm{Hz} \leq f \leq 1\,\mathrm{Hz} \quad \mathrm{goal}$$

where $S_{\mathrm{IFO}}^{1/2}$ is the effective total displacement noise in a one-way single link TM to TM measurement. This mainly concerns the interferometric measurement system, comprising the telescope, optical bench, phase measurement system, laser, clock and TDI processing. The local (TM to OB) part of that measurement has been demonstrated in LISA Pathfinder with ample performance margin, and the long arm measurement is addressed by technology development on the ground (see Section 7). That development will also benefit from the experience gained in developing the first long-distance inter-spacecraft laser interferometer on GRACE Follow-On [29], to be launched in early 2018.

*Above approximately 30 mHz* strain noise increases with frequency as the Gravitational Wave period becomes shorter than the round trip light time, resulting in a partial cancellation of the signal.

These noise levels are expected to be achievable with the following strawman parameters, to be optimised in Phase A:

- GRS from LPF with two TMs per S/C (46 mm cubic, 2 kg Au-Pt TM);
- armlength: 2.5 million km;
- telescopes with 30 cm diameter;
- laser power: 2 W end-of-life (EOL) out of the delivery fibre to the OB.

The above parameters lead to a received power of about 700 pW at the entrance aperture of the telescope, which results in a shot noise contribution of about 4.7 pm/$\sqrt{\mathrm{Hz}}$ in $S_{\mathrm{IFO}}^{1/2}$.

### 4.3 Payload conceptual design and key characteristics

A strawman design of the payload on each of the three identical S/C is illustrated in the diagram in Figure 5. It consists of two identical assemblies of roughly cylindrical shape, each of which contains a telescope, an optical bench and a GRS with enclosed TM, connected by a mounting structure which allows sequential mutual alignment during integration. The two assemblies are mounted in a common frame that allows rotation of each assembly about the vertical axis by about 2 degrees in order to track the variation of the vertex angles ($60 \pm 1°$) due to solar system dynamics.

A possible alternative configuration, which should be viewed as a backup as it would involve departures from the proven LPF GRS design, has two telescopes rigidly fixed to a single, common, optical bench and requires an "in-field pointing" actuator in each optical path to compensate the angular variation. A detailed trade-off between these options and a revised design of the payload are expected in Phase A.

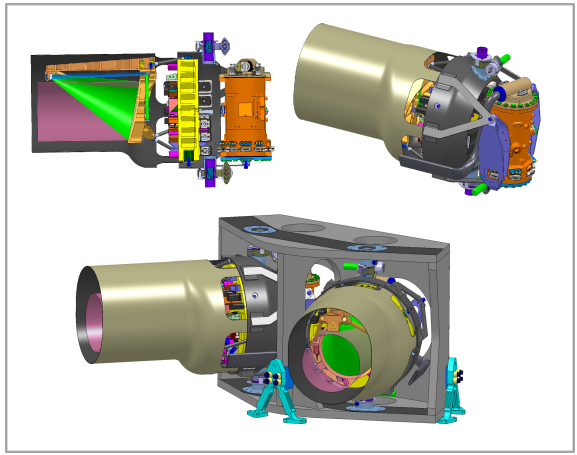

**Figure 5: Payload strawman conceptual design.** *Images courtesy of Airbus D&S GmbH, Friedrichshafen.*

### 4.4 Interferometry Measurement System (IMS)

The IMS is using optical benches which will be constructed from an ultra-low expansion glass-ceramic material to minimise optical pathlength changes due to temperature fluctuations. Each optical bench hosts one 'science' interferometer for the received light from the far spacecraft, one local interferometer which monitors the position and orientation of the test mass, and a reference interferometer. The latter two interferometers use a fraction of the two local laser beams to generate the laser beat signals. The science interferometer can use either of the two lasers together with the weak far field, to be traded in Phase A.

Construction techniques for the optical bench with the required alignment accuracy (order of 10 $\mu$m) and pathlength stability in orbit (pm/$\sqrt{\text{Hz}}$) have been demonstrated with LISA Pathfinder [4] (see Figure 6). The mechanisation of the series production of the OBs is now being studied in a technology development effort.

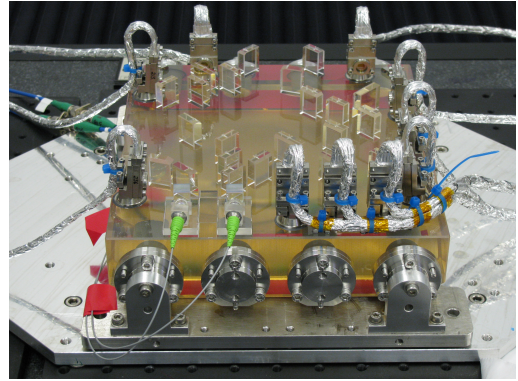

**Figure 6: The LISA Pathfinder optical bench during testing.** *Image courtesy of the University of Glasgow.*

The main laser field is injected via a single mode optical fibre and distributed via several beam splitters and mirrors to the different interferometers and additional sensors such as a power monitors. A few mW is also exchanged between the two optical benches on each S/C via the bi-directional backlink. It can be implemented via an optical fibre [30, 31], or with a free beam path between both OBs. Experimental comparisons between a few possible implementation options are ongoing at the time of writing. A possible layout of the optical bench is shown in Figure 7.

The OB has optical interfaces with the test mass on one side and the telescope on the other side. Its interface to the telescope is a precisely defined aperture (internal pupil plane) of a few mm diameter; the precise size depends on the final magnification of the telescope. Each telescope has an aperture of about 30 cm diameter and serves simultaneously the transmit (TX) and receive (RX) directions along the respective arm. In order to minimize the impact of backscattered TX light into the RX path, we assume as baseline an off-axis design with a total of about 6 curved reflectors, some of which are aspherical and which require a surface figure accuracy of about 30 nm.

An alternative is to modify the central region of the secondary mirror [32] in an on-axis design to minimise back-reflection, which would potentially simplify alignment procedures and integration. The required high stability of the optical pathlength through the telescope is expected to be achievable by the use of low-expansion materials and the very high thermal stability of the whole spacecraft [33]. If in later phases this assumption turns out to be marginal or unreliable, the envisaged design allows adding an optical truss to measure directly the phase of the outgoing wavefront without fundamental changes.

The photodiodes in the science and local interferometers are InGaAs quadrant devices with a diameter of about 2 mm with integrated preamplifiers mounted on the OB. The phasemeter processes the signals from each segment both as a sum of all segments to provide the longitudinal measurements, and differentially to provide alignment information using the Differential Wavefront Sensing (DWS) technique. DWS, used successfully in LPF and LIGO [1], measures the angle between the interfering wavefronts. The application of DWS for long inter-spacecraft links will be tested on GRACE Follow-On in early 2018 [29, 34]. This scheme provides pitch and yaw angular readouts of the TM w.r.t. to the OB, and of the S/C w.r.t. to the incoming beam, respectively. These signals will then be used as part of the DFACS [35].

Apart from mirrors, beam splitters, fibre launchers and photodiodes the OB also contains an InGaAs camera to assist in initial link acquisition, and an actuator (not shown) to compensate the slowly varying point-ahead angle, which originates from the finite light travel time along the arms (~ 8 seconds) in conjunction with the orbital motion of the spacecraft during that time.

One laser in the constellation is designated master and its frequency is stabilised to a reference cavity [36]. All other lasers are phase-locked to that master with a frequency offset in the range from 5 to 25 MHz. For redundancy and symmetry each spacecraft carries an identical cavity. Due to the time variability of the Doppler shifts, the frequencies must be switched occasionally (every few weeks) according to a frequency plan computed on ground [36]. An Electro-Optic Modulator (EOM) imprints several weak auxiliary modulations on the transmitted laser light, to transmit clock noise and allow bidirectional timestamp synchronisation, to measure absolute range distances, and to transfer data between spacecraft.

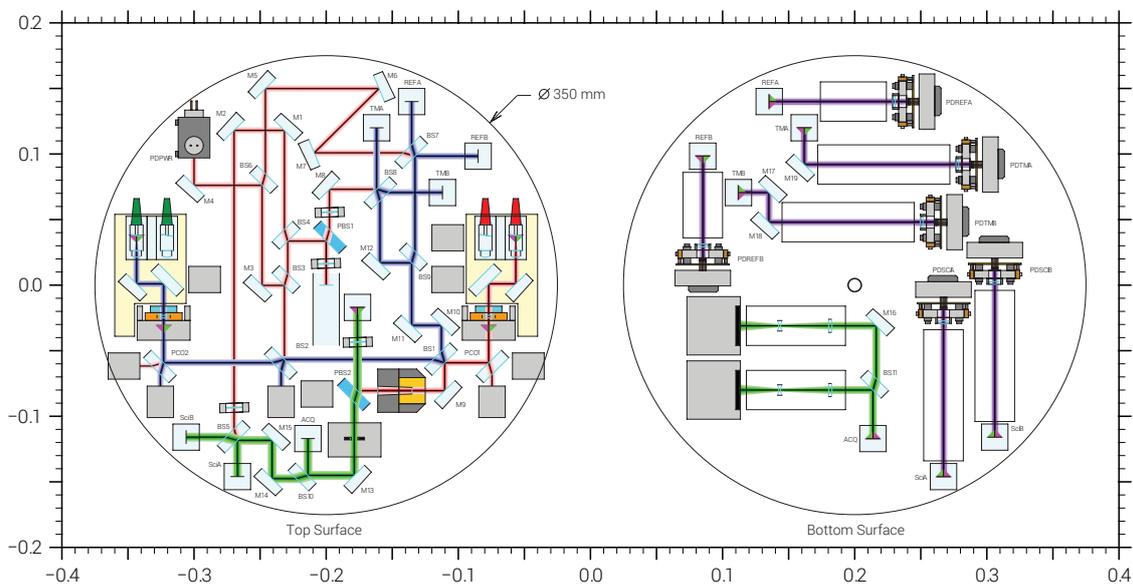

**Figure 7: Possible Layout of the optical bench for LISA.** *Image courtesy of the University of Glasgow.*

All beat note signals are processed in the phasemeter. Each channel is digitised by a fast Analogue-to-Digital Converter (ADC), for example, sampling with 14 bits at 80 MSPS, and then processed in a Field-Programmable Gate Array (FPGA). The phase and frequency of the beat note is continuously tracked by a Digital Phase Locked Loop (DPLL) with a few 10's of kHz bandwidth. Phase and frequency then exist in digital registers within the FPGA from where they can be directly extracted and decimated by digital filters. Auxiliary functions for the long-arm channels track the clock tone sidebands and the pseudo-random noise mod-ulation for use in the TDI algorithm on ground and the inter-spacecraft data transfer. Full functionality and required performance of the phasemeter have been demonstrated both in Europe and the US [37, 38].

## 4.5 Gravitational Reference Sensor

The GRS is composed of the test mass and the hardware that surrounds it. It is designed to:

- provide $z$-axis position sensing for S/C control (5 nm/Hz$^{1/2}$) level), as well as sensing used for the TM $y$ and $\theta$ control;
- provide actuation forces and torques sufficient to compensate nm/s$^2$ and 10 nrad/s$^2$ translational and angular accelerations;
- shield the TM and limit stray forces, to allow the $x$-axis free-fall requirement in Section 4.2 (roughly 3 and 12 fm/s$^2$/Hz$^{1/2}$ at, respectively, 1 mHz and 100 $\mu$Hz)

LISA requires no science-mode force actuation along the sensitive interferometer – $x$-axis, and thus there is no applied force in the LISA science signal. Electrostatic actuation is still needed, however, on all other degrees of freedom and needs to be tracked as a possible source of stray force noise.

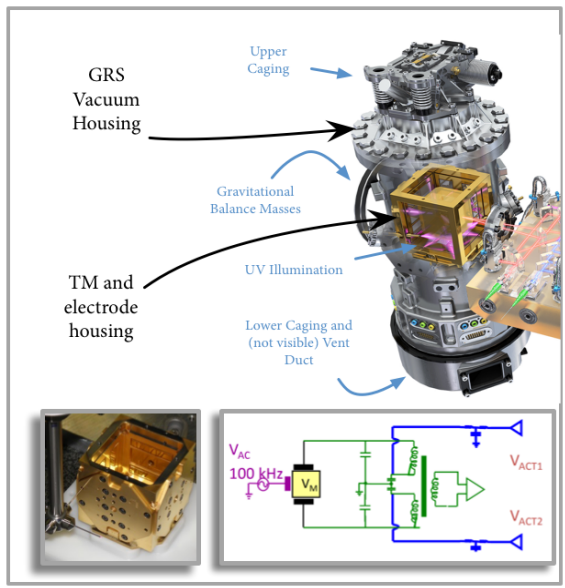

**Figure 8: Illustration of GRS hardware demonstrated on LPF.**

The GRS must provide several additional functionalities to allow this science mode performance:

- safe "caging" of the TM during launch;
- TM release and possible regrab on orbit (Grabbing Positioning Release Mechanism, GPRM);
- all-axis sensing and microNewton-level actuation to stabilize the TM during release;
- discharging to neutralise TM from cosmic ray and solar particle charging;
- TM mirror finish and line-of-sight laser beam access consistent with the pm/Hz$^{1/2}$ performance required for the local IFO readout.

Finally, in the case of either a GRS or local IFO failure, a single GRS could be used as a fallback for multiple-axis geodesic reference and control, maintaining much of the low frequency science.

The proposed GRS subsystem [39] is based on the heritage of LPF, which has demonstrated the LISA top-level acceleration noise requirement, as well as position sensing and other functionality requirements. At the GRS core is the TM itself, a 46 mm, roughly 2 kg, Au-coated cube of Au/Pt, chosen for its high density, low magnetic suceptibility, and electrostatically homogeneous and inert surface. This is surrounded, without any mechanical contact, by a similarly coated electrode housing (EH), with a 3-4 mm gap between the TM and surrounding surfaces. Electrodes on the 6 EH faces allow simultaneous 6 degree-of-freedom translational / rotational capacitive sensing and electrostatic force / torque actuation, provided by a dedicated GRS Front-End Electronics (FEE). Key design features in the GRS and FEE limit stray forces, including:

- relatively large TM-EH gaps reduce force noise from stray electrostatics and residual gas effects, which decrease with gap;
- all-AC voltage sensing and actuation limit coupling to DC and low-frequency stray electrostatic fields and TM charge variations;
- high thermal conductivity construction attenuates thermal gradients;
- nearly symmetric geometry limits cross-talk and forces acting on the different TM surfaces;
- vent-to-space vacuum chamber, to guarantee sufficiently low residual gas pressure, roughly 1 $\mu$Pa, to limit Brownian motion from molecular impacts.

Bipolar TM discharge will be performed with UV illumination and photoelectric emission from TM and EH surfaces. This has been successfully demonstrated with LPF using the 254 nm line in Hg discharge lamps. UV LEDs currently under development [40, 41] will likely allow increased flexibility, robustness, and lifetime.

The GRS is completed by auxiliary elements for measuring and mitigating various force noise sources. These include DC voltages, provided by the GRS FEE, for measuring and compensating TM charge and stray electrostatic effects, and a diagnostic system including thermometers / heaters, magnetometers / coils, and a radiation monitor.

## 4.6 Performance assessment with respect to science objectives

The LISA performance requirements, given in Section 4.2 and the corresponding strain sensitivity curve in Figure 2, are expected to deliver fully the sensitivity required in Section 3 and thus achieve the science goals of Section 2. The performance of the model payload and spacecraft described in this section is based on analysis and ground measurements (see for instance [42]) but also on LPF experimental data. The acceleration noise budget includes a number of known sources – actuation fluctuations, Brownian noise, TM charge and stray potential fluctuations, spacecraft coupling, magnetic and thermal gradient effects – with parametric models consolidated by dedicated tests on LPF which fit comfortably into the overall requirement. Key requirements needed to limit these noise sources are reflected in the system requirements in Section 5. This low frequency noise budget also includes at present an allocation for an unmodeled low frequency excess observed in LPF below 0.5 mHz (a typical LPF acceleration curve is shown in Figure 9, together with the LISA requirement). Analysis and experimentation targeted at understanding and mitigating the LPF low frequency performance is ongoing with the mission extension (through May 2017), and the complete LPF dataset will be used in the LISA Phase-0 studies to refine and consolidate the observatory low frequency performance requirements.

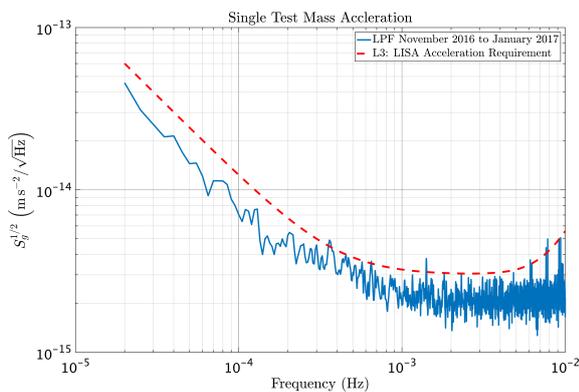

**Figure 9: Average TM acceleration noise measured with LISA Pathfinder**, compared against the LISA single TM acceleration requirement. The spectrum is the average over 12 200000 s periodograms measured in 3 separate runs between late November 2016 and early January 2017, with LPF differential acceleration noise power divided by two for comparison with the LISA single TM requirement. The data are corrected for inertial effects due to SC rotation, and roughly 10 clearly identifiable glitches have been removed from the data by fitting.

The interferometry displacement noise requirement is based on a detailed noise budget that includes not only shot noise but also allocations for pathlength variations, laser frequency and amplitude noise, clock noise, stray light, phasemeter electronics noise and tilt-to-length coupling. The local portion of the interferometry requirement has been demonstrated with ample margin by LISA Pathfinder.

## 4.7 Resources: mass, volume, power, on board data processing, data handling and telemetry

Mass, volume, power are discussed in Section 5 below since we have no clear separation of spacecraft and payload. Science data production is discussed in Section 6 below, and the data handling and downlink strategy in Section 3 above.

## 4.8 Payload control, operations and calibration requirements

In LISA, the S/C and payload work as a single entity, and this links the strategies and requirements for operations, control, and calibration. The science interferometry measurement imposes two main requirements for dynamical control during normal science operations.

First, the telescopes must be pointed to the distant S/C. With a roughly 5 microrad beam opening angle, based on a conservative analysis of achievable optical and alignment imperfections, we require a DC pointing accuracy of 10 nrad and a pointing noise below 10 nrad/$\sqrt{\text{Hz}}$. This is achieved by using the DWS angular readouts of the incoming laser wavefront as error signals that are used to guide the three-axis S/C attitude and the inter-telescope opening angle, $\alpha$. Secondly, each geodesic reference TM must be "force free" along its interferometry axis, $x_1$ or $x_2$ in Figure 10, which is achieved by drag-free control of the S/C, at the 5 nm/$\sqrt{\text{Hz}}$ level, using the local IFO measurements in the plane defined by the two interferometry axes.

The remaining S/C degree of freedom, orthogonal to the constellation plane, is drag-free controlled on the TM using capacitive sensing. All TM rotations and remaining translations – excluding the critical interferometry axes – are controlled with electrostatic actuation forces, using a combination of IFO and capacitive sensing control signals.

A calibration of the primary measurements (optical pathlength changes along the arms) is not required since these are derived from phase measurements of the MHz beat note, with the laser wavelength as the only scaling factor, and the latter can be measured on ground with sufficient accuracy. During commissioning of the payload, auxiliary calibrations will be required for pointing offsets and identifying the optimal transmit direction. These can be done by dithering one angle after the other and analysing the data on ground. Similar considerations apply for calibration and possible mitigation of known force disturbances, such as dynamic coupling to S/C motion, stray electrostatic fields, and thermal force effects, following established procedures used in LPF.

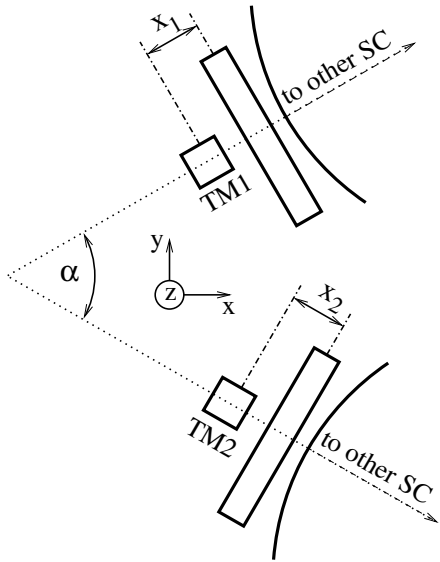

**Figure 10: Schematic of the optical arrangement within a single LISA S/C.**

Data processing on ground requires that subsystems aboard each S/C are synchronised to a single independent master Ultra-Stable Oscillator (USO), including the timing of slower processes like DFACS. This design requirement cannot be stringently derived from the top-level science requirements, but is strongly reinforced from experience gained in LISA Pathfinder.

Ideally, the science mode operation would be continuous with constant sensitivity in a very stable condition where the S/C reaches thermal equilibrium and experiences the smallest possible disturbances. However, several interruptions must be considered, all of which are consistent with near 100% duty cycle observation:

- Switching the laser phase-lock offset frequencies to keep the heterodyne beatnotes in the 5-25 MHz range, required once every several weeks. The frequency switching plan can be designed with margin to accommodate protected observation periods declared on short notice as described in Section 3;
- Antenna repointing, anticipated to be required once every 9 days, as per the communication strategy outlined in Section 3.5;
- TM discharge, to keep the charge below $10^7$ elementary charges – needed to avoid coupling to low frequency GRS potential fluctuations – could be done intermittently perhaps together with antenna repointing, or in a continuous fashion compatible with science operations. Intermittent operation may require 10's of minutes.

The frequency switching is expected to last only seconds to minutes, with negligible impact on the science data streams other than a short interruption and loss of phase continuity. Both antenna repointing and intermittent discharge are expected to be compatible with continued science mode operation, with possibly some short-interval performance degradation.

Finally, arriving into the science mode of operations will require two more operation modes:

- Constellation acquisition mode, in which 5 degrees of freedom per link (2 × 2 angles and one laser frequency) need to be simultaneously matched by using the star trackers, an auxiliary camera on the optical bench and coordinated spiral search patterns of the spacecraft attitude (see [43]).
- TM release and "accelerometer mode", with LPF heritage, where each TM is released from its "grabbed" configuration and then electrostatically forced to follow the freely orbiting S/C, before the science mode "drag-free" control is established.

# 5 System Requirements & Spacecraft Key Factors

## 5.1 System Requirements

Starting from the science requirements and the resulting strain sensitivity curve listed in Section 2, and based on the measurement principle and payload design outlined in Section 4, a strawman set of system requirements applicable to LISA has been derived. These are summarised in Table 2.

## 5.2 Spacecraft Key Factors

### 5.2.1 S/C Pointing

Of critical importance for LISA is the S/C attitude control system where a multiple degree-of-freedom con-trol system is required to control not only the attitude of the S/C and Test Masses (TMs), but also the translational degrees of freedom with a few nrad and few nm precision, respectively. Unlike in a conventional observatory, the attitude control system is intimately linked with the payload and derives its key error signals from the interferometer and the GRS capacitive sensors. The DFACS system flown on LISA Pathfinder utilised a cold gas propulsion system and achieved the required control performance. Due to this heritage, the LISA mission concept presented here utilises a similar cold gas micropropulsion system with an enlarged cold gas storage capacity to allow a maximum mission duration of 10 years. It should be noted, however, that other micro propulsion systems may also be available - in particular systems based on the colloidal thrusters tested by NASA on LISA Pathfinder or the micro-Radio frequency Ion Thruster ($\mu$RIT) investigated for the New Gravitational wave Observatory (NGO) mission [35] studied as an ESA L1 candidate would enable a mass saving compared to cold gas of around 100 kg per S/C.

### 5.2.2 Gravitational Balance and Electromagnetic Control

In order to minimise the residual acceleration, the spacecraft must be designed to minimise external forces - both gravitational and magnetic - on the test mass. In particular the spacecraft and payload design must ensure that the mass distribution is such that the residual DC torque and force on the test mass is minimised (see requirements in Table 2). This level of balancing has been demonstrated with LISA Pathfinder and so is not a new technology - it does, however, require careful design.

Similarly, there is a need to control both the absolute magnetic field and the magnetic field gradient at the GRS. Again, the capability to achieve this has been demonstrated with LISA Pathfinder and is not a new technology, but it does require careful attention to the S/C and payload design (e.g., no use of ferromagnetic materials near the GRS). It is also necessary to have a stable electrical system, with no spurious frequencies especially within the phasemeter bandwidth.

### 5.2.3 Thermal Control

In order to achieve the test mass to test mass interferometer measurement noise of $\sim 10\,\mathrm{pm}/\sqrt{\mathrm{Hz}}$, it is essential that the temperature stability of the payload within the measurement bandwidth is very high. Temperature fluctuations can couple strongly to pathlength fluctuations within the optical bench and telescope. Additionally, a very high low-frequency temperature stability within the GRS is required to reduce gas pressure fluctuations and thermal gradient effects around the TM. Very careful thermal design of the S/C is required to achieve this stability; for example the switching on/off of power consuming items on time scales which fall within the measurement bandwidth should be avoided and the S/C should remain shaded from the sun by its top face (solar panel) at all times (see Figure 11). It is also important that the absolute temperature within the core payload be within strict limits. For the Optical Bench, the operating temperature should be within ±10 K of room temperature to minimise static misalignment while it is likely that the laser system must be trimmed in orbit to be within 1 – 2 K of a setpoint to ensure stability.

### 5.2.4 Scattered Light and Contamination Control

Due to the large ratio of transmitted to received optical power within the system and the extremely high sensitivity of the read-out, control of scattered light is anticipated to be an important issue for LISA. Detailed requirements for scattered light are under study and will require better models of both the telescope and the optical bench and their interactions. Surface roughness and coating requirements for mirror surfaces appear to be well within the state of the art, but control of contamination, both particulates and thin films, may be a challenge through launch and into orbit. A detailed contamination control strategy will be formulated once the requirements are better understood.

### 5.2.5 Timing and Clock Synchronisation

Synchronisation of clocks on board each S/C is deemed to be of critical importance for the reliability of the system. In particular, all elements which fall within the primary measurement chain (e.g., phasemeter, DFACS) should derive their timing signals from a single common clock on each S/C, namely the USO of the phasemeter.

### 5.2.6 Interplanetary magnetic fields and charged particles

Fluctuations in the background magnetic field or in the flux of charged particles have a potential impact on residual acceleration of the test masses. LISA Pathfinder has provided valuable characterisation of these effects during a period of minimum solar activity and this knowledge is directly transferable to LISA since the galactic cosmic-ray spectra are known to change very little with respect distance to the Sun (3% per AU) or helio-latitude (0.3% per degree). Solar energetic particle (SEP) events connected with coronal mass ejections increase test-mass charging and can disrupt the operation of other spacecraft equipment.

While LISA Pathfinder has not experienced a major SEP event, over the course of a 4-10 year LISA mission, such events will be unavoidable and their impact on LISA science can be assessed using a combination of models verified with LPF data and in-situ measurements from other missions.

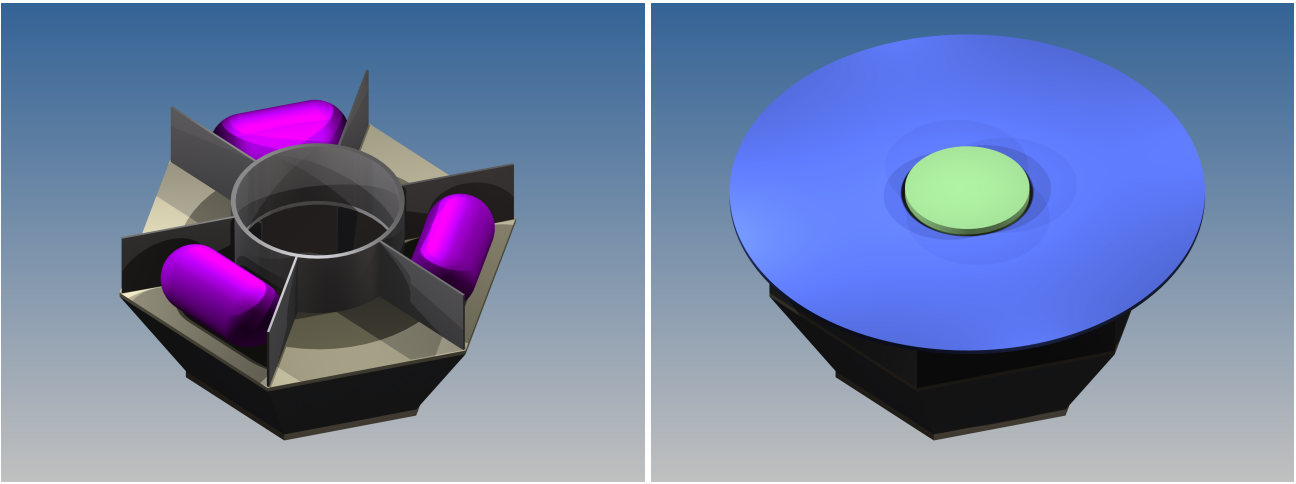

**Figure 11: (Left)** cut away showing the modified upper bay with larger cold gas tanks sized to hold sufficient propellant for a 10 year total mission duration. **(Right)** concept S/C design from the outside, showing the 2.9 m diameter flat solar array. The array is sized to ensure the S/C remains in shade at all times.

### 5.3 S/C Concept

#### 5.3.1 Structure and accommodation

Using the spacecraft design developed for the NGO mission [35] studied as an ESA L1 mission candidate as a reference, a concept for a LISA spacecraft, compatible with the payload and mission profile detailed in this proposal, has been generated. In common with NGO, we consider a modified version of the LISA Pathfinder propulsion module for LISA. Relative to the NGO S/C we propose two modifications:

1. taking advantage of the large fairing offered by Ariane 6.4, the height of the S/C payload bay is increased by 100 mm, allowing extra margin for the design and accommodation of the payload to help reduce complexity; and
2. the upper bay is enlarged to accommodate bigger cold gas storage tanks, enabling the use of a flat solar array, which is preferable.

Based on the analysis performed for NGO [35], we can estimate that the combined effect of both these modifications is to increase the mass of the S/C structure by around 5 kg for a total structure mass of 118.6 kg (including margin). In particular, we note that the central support struts for the upper floor were overspecified in NGO such that the total cold gas mass can be supported without modification. The solar array is sized to 2.9 m diameter to ensure the S/C structure remains in shade at all times and is also sufficient for power generation requirements.

#### 5.3.2 Composite Stack

The S/C model presented in Section 5.3.1, when combined with the scaled down LISA Pathfinder propulsion module investigated for NGO, has a total height of 3.4 m and a total width of 2.9 m. Applying a conservative 20% height and width margin to allow for the necessary stack adapter gives a combined stack size for three composite S/C of around 12.2 m by 3.5 m - this is compared to the Ariane 6.4 fairing which has a usable volume in the range 12.2 m by 4.3 m to 14.3 m by 3.5 m.

### 5.4 Budgets

Here we summarise the mass, power and communications budgets for the mission. These have been based on the work of the LISA Mission Formulation (LMF) [24] and NGO Reformulation [35] studies, with appropriate scaling and modifications where necessary. Margins have been applied as per ESA guidelines [44].

#### 5.4.1 Mass and Power Budget

The combined Mass and Power budget is presented in Table 6. Unit margins of 5-20% have been applied, depending on the technical maturity, as per ECSS [45]. A total system margin of 20% has been applied. For brevity, only top-level items are presented.

### 5.4.2 Communications Data Requirements and Link Budget

A data generation and communications link budget has been generated (see Table 7). The total data rate has been estimated by considering the specific needs of LISA, and where applicable using data from the LISA Mission Formulation (LMF) study and LISA Pathfinder (e.g. for housekeeping data). The driving factor in the total rate is the desired sampling rate of 3.33 Hz, which is required for a measurement bandwidth after TDI of $\leq$ 1 Hz. This data rate is for the nominal science operations mode which is used for sizing the communications system. It is envisaged that all data produced on the S/C is stored, at a high data rate, within on-board memory for a certain period of time (at least a few days). This data can be selectively downlinked for the purposes of debugging, calibration etc.

A preliminary communications link budget has been prepared to estimate the required communications band, antenna repointing frequency, achievable contact frequency and ground station requirements (see Table 3). The design is based on that proposed for NGO [35] and features a 0.5 m antenna with a 50 W transmit power in the X-Band. X-Band is preferable because it enables less-frequent re-pointing of the dish and requires less power. As per the communication strategy outlined in Section 3.5, the concept system has been designed to balance the need for daily communications with the desire to minimise antenna repointing operations (every 9 days in the present design).

### 5.4.3 LISA propulsion module sciencecraft composite $\Delta V$ Budget

Based on the use of a dedicated Ariane 6.4 launcher and injection into the reference orbit described in Section 3.1, a full $\Delta V$ breakdown has been performed. The budget is shown in Table 4 and includes margin as per [44]. The three main maneuvers are derived from the simulated orbit presented in Section 3.1. The margin for navigation and dispersion control is taken from the NGO study [35] - but is a small contributor.

### 5.4.4 Cold Gas Budget

The total amount of cold gas required for a nominal mission of 4 years plus a possible extension up to a total of 10 years has been estimated based on the consumption figures from LISA Pathfinder (see Table 5). In total, it is estimated that LISA will require 75 kg of cold gas per sciencecraft. Much of this is required to compensate the solar radiation pressure, and is thus very deterministic (and strongly coupled to the required power). There is also now extensive heritage for DFACS usage from LISA Pathfinder. As such, we apply a 20% margin to the amount of cold gas required for a total of 90 kg. A review of possible storage tanks was made, and by way of example, three Arde 5049 98.3 litre tanks would be capable of storing this volume of cold gas with a 16% volume margin. These tanks have been used in the S/C concept illustrated in Section 5.3 and the mass budget shown in Section 5.4.1.

| Parameter | Value | Driver or Justification |
|---|---|---|
| Nominal mission duration | 4 years | Duration of the main mission needed to satisfy the science case |
| Extended mission duration | 10 years | Improved parameter determination |
| Orbits | 3 heliocentric orbits | Minimise perturbations |
| Transfer time | < 18 months | Minimise time before start of operations |
| Range to Earth | 50-65 Gm | Minimise orbital perturbations without restricting communications |
| Arm length | 2.5 Gm | Resolvability of light objects |
| Number of Links | 6 links/3 arms | Polarisation sensitivity and redundancy |
| Measurement Bandwidth | Req: 100 $\mu$Hz $\leq f \leq$ 0.1 Hz<br>Goal: 20 $\mu$Hz $\leq f \leq$ 1 Hz | Detection of MBHBs at low frequencies, detection of SOBHs at high frequencies. |
| S/C Power Requirements | $\leq$ 760 W | Based on concept design, at End of Life |
| Laser Power | 2 W (out of the fiber) | Interferometer noise |
| Telescope Diameter | 30 cm | Interferometer noise |
| System wavefront quality | $\lambda$/20 RMS | Interferometer noise and jitter coupling |
| Data latency | < 1 day | Detection of light objects before merger |
| Communication Needs | 334 MB/day | 3.3 Hz data rate for 1Hz bandwidth |
| Relative timing | $\leq$ 1 ns | Required for Time Delay Interferometry |
| Absolute timing | $\leq$ 3 ns | Required for Time Delay Interferometry |
| Phase measurement bandwidth | 5-25 MHz | Phasemeter read-out noise |
| S/C jitter $\delta x, \delta y, \delta z$ | $\leq$ 5 nm/$\sqrt{\text{Hz}}$ [white] | Test mass acceleration noise and IFO cross coupling |
| S/C jitter $\delta \theta, \delta \eta, \delta \phi$ | $\leq$ 10 nrad/$\sqrt{\text{Hz}}\sqrt{1 + (3\,\text{mHz}/f)^4}$ | Minimise coupling of S/C jitter to interferometer readout |
| S/C DC mispointing | $\leq$ 10 nrad | Minimise coupling of S/C jitter |
| Temperature stability of core payload | $\leq 10^{-7}$ K/$\sqrt{\text{Hz}}\sqrt{1 + (10\,\text{mHz}/f)^4}$ | Pathlength noise for the optical bench and telescope |
| Temperature stability in GRS | $\leq 10^{-4}$ K/$\sqrt{\text{Hz}}$ at $10^{-4}$ Hz | Gas pressure noise on TM |
| Magnetic Field at GRS | $\leq$ 10 $\mu$T DC and $\leq$ 650 nT/$\sqrt{\text{Hz}}$ | Test mass acceleration noise |
| Magnetic Field Gradient at GRS | $\leq$ 5 $\mu$T/m DC and $\leq$ 250 nT/m/$\sqrt{\text{Hz}}$ | Test mass acceleration noise |
| DC S/C induced torque on TM | $\leq$ 1 nrad/s$^2$ | Test mass acceleration noise |
| DC S/C induced differential force | $\leq$ 1 nm/s$^2$ | Test mass acceleration noise |
| Charge accumulation on TM | $\leq 10^7$ e$^-$ | Test mass acceleration noise |
| Maximum pressure within GRS | $\leq$ 1 $\mu$Pa | Test mass acceleration noise |

**Table 2: Strawman System Requirements** for LISA derived from the Science Requirements.

| S/C Communication System Properties | | Ground Station Properties | |
|---|---|---|---|
| S/C Dish Diameter | 0.5 m | Atms. Loss | 3.1 dB |
| S/C Tx Frequency | 8.4 GHz | Free Space Loss [65 Mkm] | 267.2 dB |
| 3dB Beam Width | 5° | Signal at GS | -224.08 dBW |
| S/C Antenna Gain | 30.3 dBi | GS Dish Diameter | 35 m |
| S/C Tx Power | 50 W | GS G/T | 52 dB/K |
| S/C Line Losses | 1 dB | GS Losses | 1.1 dB |
| S/C Downlink EIRP | 46.22 dBW | C/N0 | 55.44 dBHz |
| | | Symbol Rate | 279 kbps |
| | | Gain Margin | 3 dB |
| Achievable User Data Rate [Turbo code] | | | |
| Data Rate [Perfect Pointing] | | 139 kbps | |
| Data Rate [±1.5 days offset] | | 108.5 kbps | |

Table 3: LISA Communications Link Budget.

| Transfer Properties | | | |
|---|---|---|---|
| Launch Date | 18.12.2030 | | |
| Arrival Date | 16.01.2032 | | |
| Breakdown | | | |
| Maneuver | SC1 [m/s] | SC2 [m/s] | SC3 [m/s] |
| Post-Launch | 204.1 | 228.5 | 76.4 |
| Inclination | 147.1 | 407.2 | 433.7 |
| Stopping | 599.9 | 632.9 | 681.3 |
| Navigation and Dispersion Control | 73.5 | 73.5 | 73.5 |
| Total ΔV per S/C | 1024.6 | 1342.1 | 1264.9 |
| Fuel Mass [kg] | | | |
| Fuel Mass at Isp = 270 s | 555.1 | 775.5 | 719.5 |
| Total Fuel Mass [kg] | 2050.2 | | |

Table 4: LISA propulsion module science-craft composite $\Delta V$ Budget for Transfer.

| Parameter | Value | Comment |
|---|---|---|
| Scaling Factor | 1.53 | Ratio of LISA and LPF solar array areas |
| Mean Thrust per DOF | 16.2 $\mu$N | Scaled from LPF |
| DFACS Consumption | 20.3 g/day | Scaled from LPF usage of 10 g/day |
| Mission Duration | 10 years | Extended Mission |
| DFACS Cold Gas Mass | 73.98 kg | |
| Maneuvers Cold Gas Mass | 1 kg | De-spin, antenna rotation etc. From NGO |
| Total CG Mass | 75 kg | |
| Total CG Mass with margin | 90 kg | With 20% Margin |

Table 5: LISA cold gas mass budget.

| Item | Mass [kg] | Power [W] | Comment |
|---|---|---|---|
| Ariane 6.4 Launch Capacity | 7000.0 | | To $V_\infty \sim 260$ m/s |
| **Total Launch Mass** | **6076.3** | | **Wet mass and adapter** |
| Stack and Launch Adapter | 500.0 | | Estimate from ESA TN |
| Wet Stack Mass | 5576.3 | | Composite S/C dry ×3, plus total propellant |
| Total Propellent | 2050.0 | | Combined for 3 S/C |
| Total Power Available | | | |
| Available Power at EOL | | 780.6 | 2.9 m array, 128 W/m² after 10 years at 30° to sun |
| Breakdown | | | |
| Composite S/C (Dry) | 1175.4 | 760.8 | Sciencecraft and Propulsion Module [+ 20% system margin] |
| Sciencecraft | 768.7 | 637.5 | Bus + Payload |
| Bus | 471.1 | 354.7 | Sum of entries below. |
| AOCS | 199.1 | 116.3 | Including 90 kg Cold Gas and 58 kg tanks |
| COMS | 33.3 | 128.2 | Based on NGO |
| On-Board Computer | 17.2 | 39.7 | Based on NGO |
| Power Subsystem | 58.2 | 10.5 | Scaled from NGO for revised solar array |
| Thermal Control | 15.8 | 60.0 | Based on NGO |
| Structure | 118.6 | | Scaled from NGO with adapted S/C model |
| Harness | 28.9 | | Based on NGO |
| Payload | 297.6 | 282.8 | Sum of entries below. |
| Structure | 44.4 | | Scaled from NGO |
| GRS | 19.7 | | LPF Flight Mass [2 in payload] |
| Telescope | 9.3 | | Estimate for 30cm Telescope [2 in payload] |
| OB | 17.5 | | Estimate for 40 cm OB [2 in payload] |
| Electronics | 54.0 | 85.4 | Including GRS Electronics and CMS |
| Laser Systems | 33.6 | 101.0 | Based on NGO, contains both Tx lasers |
| Phasemeter | 24.0 | 45.6 | Scaled from NGO |
| Diagnostics | 1.8 | 14.4 | Based on NGO |
| Payload Computer | 9.6 | 36.4 | Based on NGO |
| Payload Harness | 37.2 | | Based on NGO |
| Propulsion Module | 210.8 | | Sum of entries below. |
| Structure | 89.9 | | Based on NGO/LPF |
| Separation Sys. | 10.1 | | Based on NGO/LPF |
| Thermal Control | 11.3 | | Based on NGO/LPF |
| AOCS | 91.1 | | Based on NGO/LPF |
| Harness | 8.4 | | Based on NGO/LPF |

Table 6: Combined Mass and Power budget for LISA.

| Source | Class | Measurement | Count | Sampling Rate [Hz] | Bits / channel | Rate [bits/s] |
|---|---|---|---|---|---|---|
| Payload | | | | | | |
| Phasemeter | IFO Longitudinal | Science IFO | 2 | 3.3 | 32 | 213.3 |
| | | Test Mass IFO | 2 | 3.3 | 32 | 213.3 |
| | | Reference IFO | 2 | 3.3 | 32 | 213.3 |
| | | Clock Sidebands | 2 | 3.3 | 32 | 213.3 |
| | IFO Angular | S/C $\theta,\eta$ | 4 | 3.3 | 32 | 426.6 |
| | | TM $\theta,\eta$ | 4 | 3.3 | 32 | 426.6 |
| | Anciliary | Time Semaphores | 2 | 3.3 | 96 | 639.9 |
| | Optical Monitoring | PAAM Longitudinal | 2 | 3.3 | 32 | 213.3 |
| | | PAAM Angular | 4 | 3.3 | 32 | 426.6 |
| | | Optical Truss | 6 | 3.3 | 32 | 639.9 |
| GRS FEE | GRS Cap. Sensing | TM $x,y,z$ | 6 | 3.3 | 24 | 480.0 |
| | | TM $\theta,\eta,\phi$ | 6 | 3.3 | 24 | 480.0 |
| Payload Computer | DFACS | TM applied torques | 6 | 3.3 | 24 | 480.0 |
| | | TM applied forces | 6 | 3.3 | 24 | 480.0 |
| | | S/C applied torques | 3 | 3.3 | 24 | 240.0 |
| | | S/C applied forces | 3 | 3.3 | 24 | 240.0 |
| | Payload HK | e.g. Temperature, Power Monitors *etc.* | | | | 2613 |
| | Total Payload | | | | | 8639 |
| Platform | | | | | | |
| | Housekeeping (based on LPF) | | | | | 1189 |
| | Total Platform | | | | | 1189 |
| Totals | | | | | | |
| | Raw rate per S/C | | | | | 9828 |
| | Paketisation overhead [10%] | | | | | 983 |
| | Packaged rate per S/C | | | | | 10811 |
| | Packaged rate for Constellation | | | | | 32433 |

Table 7: LISA Data Generation Rate Breakdown.

# 6 Science Operations and Archiving

LISA science operations are envisioned to be a joint effort between ESA, NASA, and the LISA Consortium, sharing responsibility for producing and validating the various science products of the mission, participating in all levels of science operations planning, as well as populating and maintaining science archives, both during operations and in the longer term.

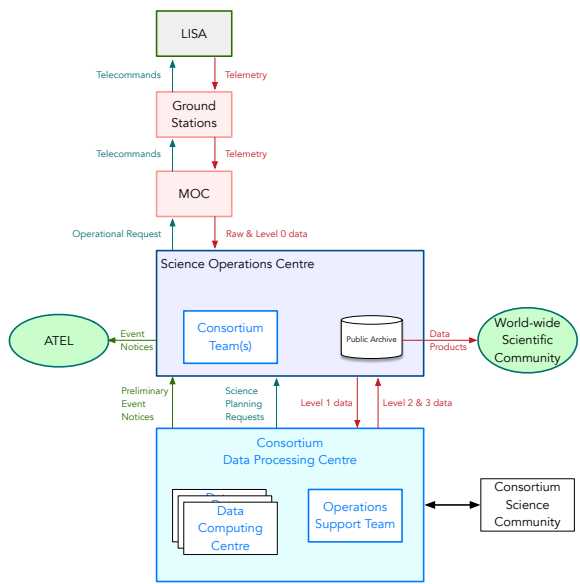

**Figure 12: A schematic of data and information flows between the different mission elements.**

Science operations, data processing, dissemination and archiving will follow existing ESA and NASA standard models, with work in the Science Operations Centre (SOC) shared by ESA and Consortium personnel and in the Consortium Data Processing Centre (DPC). To ensure tight coordination in science planning and observatory health monitoring, the Consortium will install operations teams within the SOC. This approach was taken for LISA Pathfinder science operations, where daily close contact between scientists, Mission Operations Centre (MOC), and Science Technology Operations Centre (STOC) proved extremely effective. The SOC will interface with the unique Consortium DPC, that will direct and supervise the data analysis and processing activities of the Consortium, leveraging the Data Computing Centres (CPUs and storage) provided by member states, as well as possibly ESA and NASA.

During science operations, the SOC supported by Consortium-provided operations teams where appropriate and the Consortium DPC, will perform a range of activities:

*Observatory operations* The SOC will be responsible for all observatory operations activities.

- In-flight calibration, and calibration monitoring throughout operations;
- Maintaining up-to-date calibration files to be used in data processing;
- Monitoring operations and triggering configuration

updates.

*Data pre-processing (at the SOC)* We expect routine data volumes of about 300 Mbytes/day, L0 data, arriving at the ground stations yielding about 600 Mbytes/day of Level 1 data. In this context, Level 0 data refers to raw science telemetry and housekeeping data; Level 1 data to TDI variables, all calibrated science data streams, and auxiliary data. The SOC will be responsible for generating Level 1 data products. The SOC activities then include:

- Ingestion of Level 0 data from the MOC;
- Rapid troubleshooting with a quick-look analysis of Level 0 data;
- The preprocessing and calibration of raw telemetry;
- Producing Level 1 data products (TDI variables) using Consortium-provided data-processing pipelines (e.g., generating TDI observables and data-quality flags);
- Transfering Level 1 data to the Data Processing Centre.

*Data Analysis (by the DPC)* The data analysis for LISA, which is the responsibility of the DPC, covers the identification and extraction of signal waveforms and the composition of source catalogues. The Consortium DPC will deliver Level 2 data to the SOC, which corresponds to intermediate waveform products such as partially regressed observable series (i.e., a dataset obtained by progressively deeper subtraction of identified signals), as well as Level 3 data which corresponds to catalogues of identified sources, with faithful representations of posterior parameter distributions. Level 2 and Level 3 data represents approximately 6 Gbytes/day.

The Consortium DPC will generate and distribute the main LISA science products to the SOC and to the Consortium science community. The periodic release to the community at large (per relevant ESA and NASA data-release policies) is under the responsibility of ESA.

Two Data Computing Centres (DCCs), one in Europe and one in the US, are planned as providing the local computing hardware, but it is highly likely that more will exist in Europe, in particular in Germany, UK, Spain, and Italy. National funds are assumed to develop and operate the European Data Computing Centres (DCCs). CNES has performed a Phase 0 study on the LISA DPC [46], and will be funding, in collaboration with other participating countries, the development of the Consortium DPC for central coordination of data analysis at all DCCs. The DPC activities include:

- Receiving Level 1 data from the SOC;
- Creating Level 2 and Level 3 science products;
- Analyzing the quality of science data products;
- Producing periodic science data product releases;
- Generating alerts for upcoming transient events, such as mergers;

*Transient events processing* Part of the Level 2 data includes rapid notification of transient events for the astronomy community. The requirement to provide these notifications routinely within a day or so of observation will set the general latency requirement on Science Operations elements. Preliminary transient event notices received from the Consortium DPC must be assessed for quality and then prepared for publication. This includes:

- Producing and assessing preliminary event notices;
- Using established channels to notify astronomical community;
- Providing detailed transient parameters to the science planning team.

As an all-sky observatory, LISA detects multiple (potentially 1000's), overlapping signals throughout the operational period, and identifying and extracting these signals from the data stream is a challenging data analysis task. In anticipation, a series of highly successful Mock LISA Data Challenges engaged the scientific community for a number of years, and it is anticipated that similar activities will take place in the future to help drive and coordinate the development of the necessary data analysis pipelines and signal waveforms. Although some data processing and waveform generation details remain, the data analysis is considered to be tractable.

*Science planning* The SOC is the unique point of contact with the MOC to facilitate the payload uplink chain:

- Planning observatory operations requests to update configurations;
- Planning calibration activities;
- Scheduling special observing periods.

Figure 12 shows a schematic of the role of LISA science operations within the LISA mission management structure and highlights key data and information flows.

*Protected Observation Periods* LISA science requirements dictate special observing periods during which no observatory maintenance activities should take place, allowing the observatory to remain in its nominal science mode. Such periods will be triggered in advance according to predictions made on, for exam-ple, the merger of a MBHB, and will be fed into the science planning

*Archive data quality* The LISA Consortium anticipates at least two complete and identical archives (one in Europe and one in the US, for example). It is important to ensure that these archives provide consistent higher level products to the scientific community. The high level products generated by the DPC will be consolidated and validated by the Consortium before they are issued to the archives.

*Data Release Policy* Data releases will follow the general practices of ESA and NASA science missions. Publication of updated Level 3 science products is expected to occur periodically, for example, after each 6-months to 1-year of data collection. Some data products, such as transient notifications, will be released in near real-time upon processing. After an agreed upon proprietary period, Level 1 and Level 2 data will made publically available.

# 7 Technology Development Requirements

Unlike most mission concepts in the pre-project phase, the LISA mission concept enjoys a high level of technical readiness. This is due in large part to the significant international effort on the LISA Pathfinder mission, technology development efforts in Europe and the US associated with the LISA project since the early 1990s, and work on related missions such as the Laser Ranging Instrument on GRACE-FO and Gaia. In Table 8 below, we briefly summarize the current level of readiness of the key technologies for our proposed concept. The items are color-coded according to their status with items in green requiring little or no development, items in yellow requiring some development, and items in blue representing desirable technologies which would provide a specific benefit such as reduced mass, increased science performance, or increased margin on specific error budget items, but are not required to meet the performance or mission requirements outlined in this proposal.

Some of the technology items have been tested on LISA Pathfinder and are therefore at TRL 9. For some of them, like the test-mass release mechanism operation sequence, some optimization is nevertheless desirable.

## 7.1 Algorithms/ methodology / simulation / data processing

The analysis and mitigation of possible straylight effects needs some attention. Straylight includes scattered light from surface imperfections, dust etc., but also ghost beams originating from spurious reflections at nominally transmissive glass interfaces. Available commercial tools do not normally provide the required answers. Therefore, an intensive effort is required to improve the modeling and simulation tools, to verify simulated results in experiments that emphasize specific aspects of the simulation, and to start early to build optically representative models of the optical payload and investigate its stray light related behavior.

Data preprocessing to recover arm lengths and clock offsets, and TDI processing has been under study at several places, but needs further development.

Technology development activities including those listed in the table are currently funded and organized independently by ESA, numerous Consortium member states, and NASA. As the mission matures into the project phase, it is expected that these activities will be brought under the control of an ESA project office which will in turn coordinate any activities funded by the Consortium or international partners. Equally important is an exploration and timely resolution of potential design trades that can have significant ripple effects (in many cases beneficial) on the rest of the system. The most prominent example is the trade between an articulated fixed-mirror telescope with a narrow field-of-view and a fixed, wide field-of-view telescope with a moving mirror to track the far spacecraft. It is expected that the examination and resolution of such trades will be an important element of the early industrial system study phase.

Based on LPF heritage and ongoing technology development, we trust that all technologies can be at least at TRL 6 by 2020.

| Technology | Status | TRL |
|---|---|---|
| Gravitational Reference Sensor Technologies | | |
| Test mass electrostatic readout and actuation | On-orbit LPF performance used to develop sensitivity curve. Some flexibility allowed in Phase-A. | 9 |
| Caging and release mechanism | Launch-lock, release, and re-grabbing functions demonstrated on LPF. | 9 |
| Charge Management System | UV Charge control demonstrated on LPF. | 9 |
| - UV Source: Hg lamps | LPF Heritage (lifetime to be investigated). | 6 |
| - UV Source: LEDs | Development efforts in UK and US. Charge control demonstrated on torsion pendulum. | 4 |
| Drag-free Attitude and Control System (DFACS) Technologies | | |
| DFACS control algorithms | 18 DoF control demonstrated on LPF by both DFACS (ESA) and DRS (NASA) algorithms, performance meets LISA specs, LISA version will add constellation pointing requirements, performance simulated in prior LISA studies. | 7 |
| Cold Gas Micropropulsion | Thrust noise requirement demonstrated on LPF. Additional heritage from GAIA and Microscope. | 9 |
| Colloidal Micropropulsion | Thrust noise requirement demonstrated on LPF. Additional development required for redundancy and lifetime. | 7 (head), 5 (feed system) |
| miniRIT & HEMP Micropropulsion | Laboratory work ongoing | 4 & 3 |
| Laser System Technologies | | |
| Master Oscillator - TESAT NPRO | Full heritage (TESAT) on LPF and GRACE-FO. All requirements met. | 9 |
| Fiber Amplifier - TESAT | Significant flight heritage at required power levels (NFIRE, TerraSAR, AlphaSat, Sentinel). Laboratory campaign to verify phase fidelity underway (CFI component). | 5 |
| Fiber Amplifier | Ongoing development effort at GSFC. Meets noise requirements including sideband stability. Partial environmental testing done. 2.5 W output power (CFI component). | 4 |
| Frequency Reference Cavity | Flight Optical cavities for GRACE-FO delivered and demonstrated in laboratory (US) to meet all LISA requirements. Equiv. European development ongoing | 8 |
| Master Oscillator - ECL | Ongoing development effort at GSFC in partnership with US industry. | 4 |
| Optical Bench Technologies | | |
| Bonding Technology | Alignment stability and displacement noise requirements demonstrated on-orbit with LPF. | 9 |
| Fibre injectors | Pointing stability and beam quality requirements demonstrated on-orbit with LPF. Prototype for LISA has been raised to bread-board level. | 5 |
| Manufacturing | Efforts underway (UKSA & ESA funded) to optimize manufacture process to reduce construction time and schedule risk. | 4 |
| Photoreceivers - US | Two parallel efforts at JPL and GSFC in partnership with US industry. Laboratory prototypes demonstrate improved noise performance. | 4-5 |
| Photoreceivers - DLR/Adlershof | Heritage from GRACE-FO (TRL 8), requires moderate performance improvements. | 4 |
| Interferometric phase reference | Several variants studied in laboratory environment. Design and testing consolidated under ESA-funded activity. | 4 |
| Pointing Mechanisms | Two prototype Point-Ahead Angle (PAA) mechanisms developed (TNO & RUAG) and tested in a laboratory environment. | 4 |
| Telescope Technologies | | |
| Optomechanical Stability | Pathlength stability of a representative metering structure demonstrated in laboratory. | 4 |
| Optical Truss | Risk mitigation against insufficient optomechanical stability. Some heritage from GAIA but requires adaptation to LISA requirements. | 4 |
| Pointing - Articulated Telescope | Four-optic fixed mirror design developed and prototyped. Candidate articulation actuator noise performance validated in NASA laboratory study. | 4 |
| Pointing - In-field Guiding | Optical design completed and prototyped (Airbus DS), including candidate optical bench interface. | 3 |
| Phase Measurement System Technologies | | |
| Complete functionality | German / Danish Phasemeter from ESA CTP and JPL lab work | 4 |
| Core functionality | JPL Phase measurement, DWS angle sensing, closed-loop laser frequency control demonstrated on GRACE-FO flight units. | 8 |
| LISA-specific functions | clock transfer, jitter calibration, and ranging demonstrated in laboratory prototypes. | 4 |
| Diagnostics Technologies | | |
| Diagnostic Items | LPF Heritage (TRL 9), LISA adaption for the temperature and magnetometer sensors to be done. | 4 |

Table 8: **Technology readiness levels of primary mission items.**

# 8 Management Scheme and Cost Analysis

The LISA proposal in response to the Call for L3 mission concepts is submitted by an international collaboration of scientists called the LISA Consortium. Our proposal is fully compliant with the science goals indicated in the "Report of the Senior Survey Committee on the selection of the science themes for the L2 and L3 launch opportunities in the Cosmic Vision Programme" [2]. The team is building on the proto-consortium that proposed a Gravitational Wave observatory for the L1 flight opportunity, but has been growing considerably. It is augmented by additional member states and the US as international partner. The LISA Consortium also proposed *The Gravitational Universe* [3] as a science theme for the selection of the L2 and L3 launch opportunity and submitted the pertinent White Paper. The LISA Consortium also comprises all the investigators who have successfully pursued the LISA Pathfinder mission, a number of scientists who worked on the ground-based LIGO, Virgo, and GEO projects, and the Laser Ranging Interferometer on the GRACE Follow-On mission, thus making full use of all the expertise that has accumulated. This approach optimises the utilisation of the remaining time for mission preparation and technology development. We expect all mission elements to be at least at TRL 6 around 2020.

Recognizing that on LISA all aspects of the mission performance are inseparable and have to be studied and coordinated centrally from the three-satellite constellation down to rather low level in the payload, we recommend a strict top-down approach, with ESA appointing a single prime contractor to take responsibility for the payload as well as the S/C. For the same reasons we also foresee ESA setting up a System Engineering Office for the entire mission including the core scientific instrumentation, providing top-level overview of all mission aspects including the payload. This office gives close guidance to the Payload Coordination Team provided by the Consortium. The Consortium will support the ESA System Engineering Office with key personnel providing expert knowledge on the critical aspect of the detector, including that gathered from LISA Pathfinder, as requested by ESA.

The Consortium will also deliver to ESA the integrated science instrument at the heart of the payload, plus several spacecraft-mounted parts of the instrument. It is expected that the remaining parts of the payload, in particular lasers and telescopes, will be procured by ESA or provided by NASA.

The LISA Science Instrument will comprise the optical bench with attached GRS and detached Phasemeter and Data and Diagnostics Subsystem (DDS) plus the supporting electronics, comprising the GRS Front-end Electronics, Charge Management System, Caging system, and Data Management Unit (DMU). The instrument will be funded by the European member states, with likely contributions from NASA.

The distribution of tasks in the Consortium and the commitments to deliver various pieces of flight hardware, ground support equipment, data processing, and integration tasks will closely follow the preliminary declarations of intent brought forward by the member states during the past meetings of the ESA-appointed Gravitational Wave Observatory Working Group (GW-WG).

Germany has the lead role in the Consortium and provides the Consortium management, System Engineering for the instrument under the lead of the ESA Mission System Engineering office, and the Phasemeter (PM) system, including USO and Frequency Distribution System, with possible contributions from the Netherlands, Belgium, Switzerland, Denmark and/or NASA. France takes the responsibility for integration and performance control of the Science Instrument, and provides System Engineering support to Germany on these aspects. France participates in the provision of the Instrument mounting structure with a detailed contribution depending on the results and the recommended design from the forthcoming Phase A. France also provides the DPC for the mission. Italy will be responsible for delivering the integrated and tested Gravitational Reference Sensor, that includes the Test Mass, the Electrode Housing, the Vacuum Container, the Gravitational Balance Masses, the Charge Management System, the Caging Mechanism (Swiss contribution), and the Front-End Electronics (Swiss contribution). Italy will also support Germany with system engineering aspects that involve GRS and acceleration disturbance aspects. The UK provides the assembled optical benches, and will provide System Engineering support to Germany on aspects involving the metrology system performance and payload alignment. Switzerland provides the Front-End Electronics and the Caging Mechanism for the GRS. Spain provides the DDS including the DMU and electronics. Denmark provides contributions to the Phasemeter system. Belgium provides the acquisition sensors on the optical benches and the photoreceivers comprising photodiodes and front ends. The Netherlands provide electronics for the photoreceivers and possibly the actuators on the optical benches. Sweden will lend support to the data processing and analysis. Portugal will provide elements of the data analysis and the qualification of optical and electronic components. Hungary will provide elements of the data analysis.

The Consortium is proposing that NASA provide directly to ESA further parts of the payload and the spacecraft. There may also be NASA contributions to the Consortium. The total NASA contribution is expected to be at a level of 20% of the total mission cost. All elements proposed to be provided by our international partner have potential back-up solutions based on European technology.

The Consortium will deliver the integrated, tested and aligned LISA instrument to the ESA P/L contractor, who will in turn integrate the Instrument and the ESA-procured or NASA-provided telescope and laser to complete the payload.

## 8.1 LISA Consortium Organisation, Roles and Responsibilities

### 8.1.1 LISA Consortium

The LISA Consortium is based on 12 European countries and the US, currently representing more than 300 scientists as Consortium members (https://www.lisamission.org/consortium/) and more than 1300 researchers as supporters (https://www.lisamission.org/supporters/). The Consortium Board comprises one or two representatives from each country as the Co-Investigators. It is the final decision making body and is responsible for the implementation in the participating countries. Top-level operations are coordinated by the Executive Board which is led by the Consortium Lead as the single point of contact with ESA.

### 8.1.2 Top Level Organisation in Phase A/B

Figure 13 illustrates the top-level organisation of the LISA Consortium and project during the Phase-A/B of the project. The organisation is optimized to support the study activities during this mission phase and overseeing the technology development activities.

The LISA Consortium is led by the PI, called LISA Consortium Lead (LCL), supported in its top management function by the Executive Board consisting of the Consortium Lead (Karsten Danzmann) and five Co-PIs, one for Science (Pierre Binetruy), one for LISA Pathfinder (Stefano Vitale), one for the optical metrology (Henry Ward), one as liaison to the other member states (Domenico Giardini), and one as liaison to the US community (David Shoemaker). The composition of the Executive Board will be adapted to the changing project needs. The LISA Consortium Lead is the single formal interface of the Consortium with ESA

The scientific community is represented in the LISA Consortium Board, comprising Co-Investigators as representatives per participating country. The Consortium Board is the final decision making body and meets at least four times per year.

The Consortium Lead receives scientific supervision and advice on scientific matters from the Science Study Team comprising scientists from the international scientific community, appointed by ESA and chaired by the ESA Project Scientist. The ESA Science Study Team is responsible for the definition of the mission scientific requirements that are recorded within the Science Requirement Document (ScRD) and for ensuring that the implementation of S/C, P/L, instrument, and units of LISA fulfills these scientific requirements.

The Consortium Lead is supported in this management function by the Payload Coordination Team, the Ground Segment Coordination Team, and the Science Coordination Team, each led by a Team Coordinator.

Close guidance to the Payload Coordination Team is given by the ESA Systems Engineering Office. The Systems Engineering Office is an ESA function providing top-level overview of all mission aspects including the payload, recognizing that on LISA all aspects of the mission performance are inseparable and have to be studied and coordinated centrally down to rather low level in the payload. The Consortium will provide members to the Systems Engineering Office as requested by ESA. NASA will also contribute to the Systems Engineering.

The ESA Steering Committee is the representation of the participating funding agencies of the ESA Member States and NASA, and responsible for the final decision about financial implications at the national level.

**LISA Consortium Board**

The role of the LISA Consortium Board is to define the Consortium policy with respect to the Consortium management and the scientific objectives. The Consortium Board steers the activities of the Consortium in the involved countries, it confirms the members of the Executive Board in agreement with ESA and the Steering Committee, and delegates the management and the coordination of the Consortium and the top-level operative decisions to the LISA Consortium Lead and the Executive Board. The Consortium Board also defines the topics and appoints the conveners of the Working Groups, after consultation with the Executive Board and the Science Coordination Team. The Consortium Board meets at least four times per year.

The Consortium Board comprises Co-Investigators as representatives per participating country and is chaired by the Consortium Lead. The current members of the Consortium Board are Karsten Danzmann and Bernard Schutz (D), Stefano Vitale and Monica Colpi (I), Pierre Binetruy and Nary Man (F), Domenico Giardini and Philippe Jetzer (CH), Harry Ward and Alberto Vecchio (UK), Carlos Sopuerta (ES), Allan Hornstrup (DK), Gijs Nelemans (NL), Thomas Hertog (B), Vitor Cardoso (PT), Ross Church (SE), Zsolt Frei (HU), David Shoemaker, Neil Cornish, Guido Mueller and Shane Larson (US). The membership may be changed to adapt to the evolving project needs.

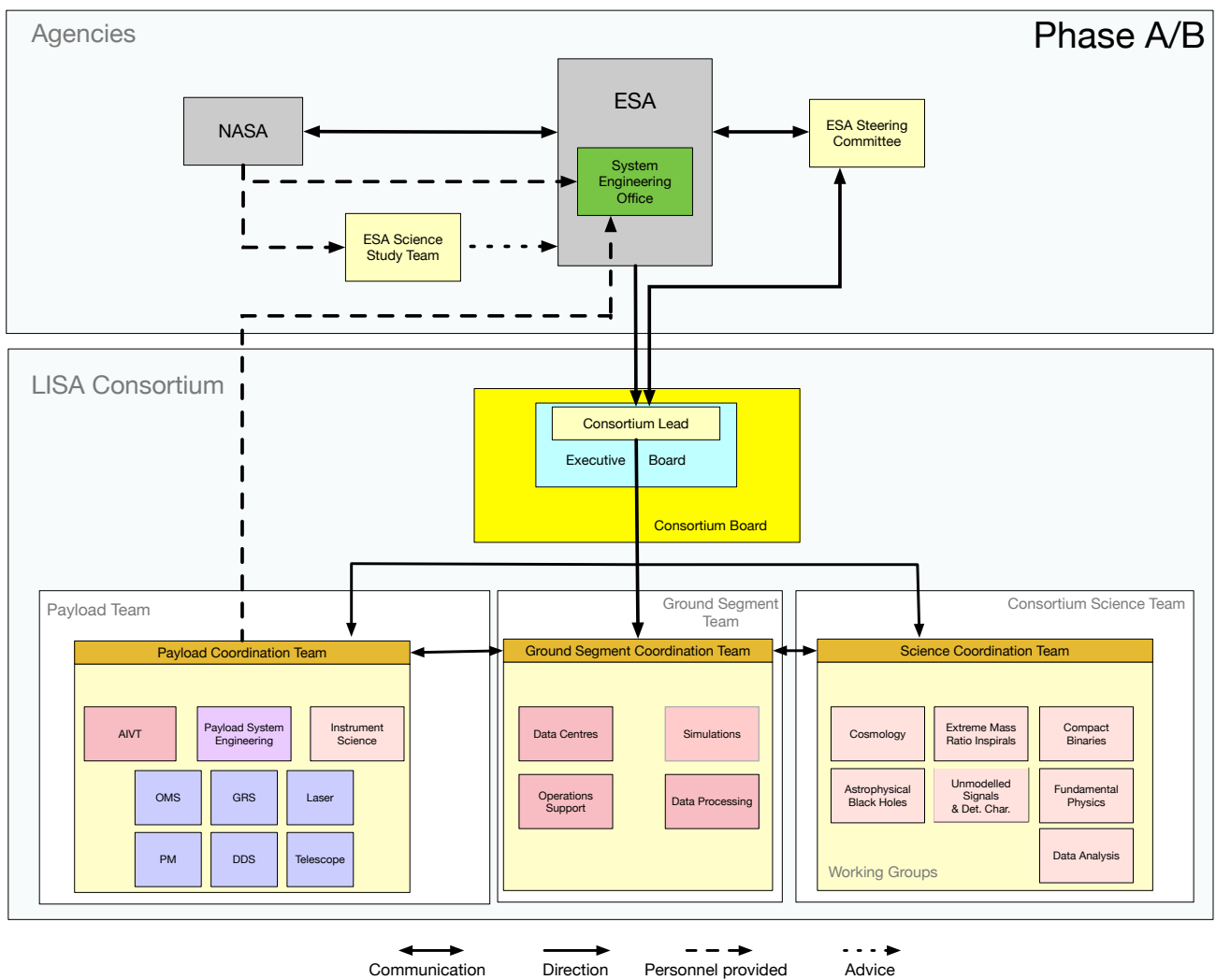

**Figure 13: LISA Organisation in Phase A/B.** OMS: Optical Measurement System, PM: Phasemeter, CMU: Charge Management Unit; DDS: Payload Computer and Diagnostics.

**Payload Coordination Team**

The role of the Payload Coordination Team is to support the Phase-A study by closely following the study of all payload units, AIVT, and SE; as well as to coordinate the technology development of all units. The Payload Coordination Team comprises the AIVT, Payload SE, and Instrument Science Study Leads, and the Study Leads for each of the Payload Units, and it is coordinated by the Payload Coordination Team Lead.

**Ground Segment Coordination Team**

The role of the Ground Segment Coordination Team is to support the Phase-A study in all aspects of the ground segment and data processing. It comprises the Study Leads for Simulations, Data Processing, Operations Support, and Data Centers, and is coordinated by the Ground Segment Coordination Team Lead.

**Science Coordination Team**

The role of the Science Coordination Team is to support the Phase-A study in all aspects of the astrophysics and the sources to be addressed by the LISA mission. It identifies fixed term projects, and their project leaders, in order to fulfill these needs. The projects are transverse to the Science Working Groups. The members of the Science Coordination Team are appointed by the Consortium Board on recommendation by the Executive Board. They can be chairs of Working Groups, but do not have to be. The Science Coordination Team is chaired by the Science Coordination Team Lead.

**Science working groups**

The role of the science working groups is to allow the LISA community at large to focus on the scientific objectives of the mission. Currently there are 7 science working groups, but the number may change as our understanding of the science develops. The current working groups are Cosmology, Extreme Mass Ratio Inspirals, Compact Binaries, Astrophysical Black Holes, Unmodeled Signals and Detector Characterisation, Fundamental Physics, and Data Analysis.

The exact list of working groups, as well as the names of their chairs, is formally decided by the Consortium Board after consultation with the Executive Board and the Science Coordination Team.

**Organisation during Phase C/D**

For Phase C/D the project organisation has to undergo considerable changes, as the responsibility for studies transforms into the responsibility for hardware delivery on time and on schedule. The former Study Leads will be replaced by managers with clear executive responsibilities. The Coordination Teams will be transformed into Management Teams with responsibility for schedule and budget. The Unit Study Leads will become Instrument Unit Managers.

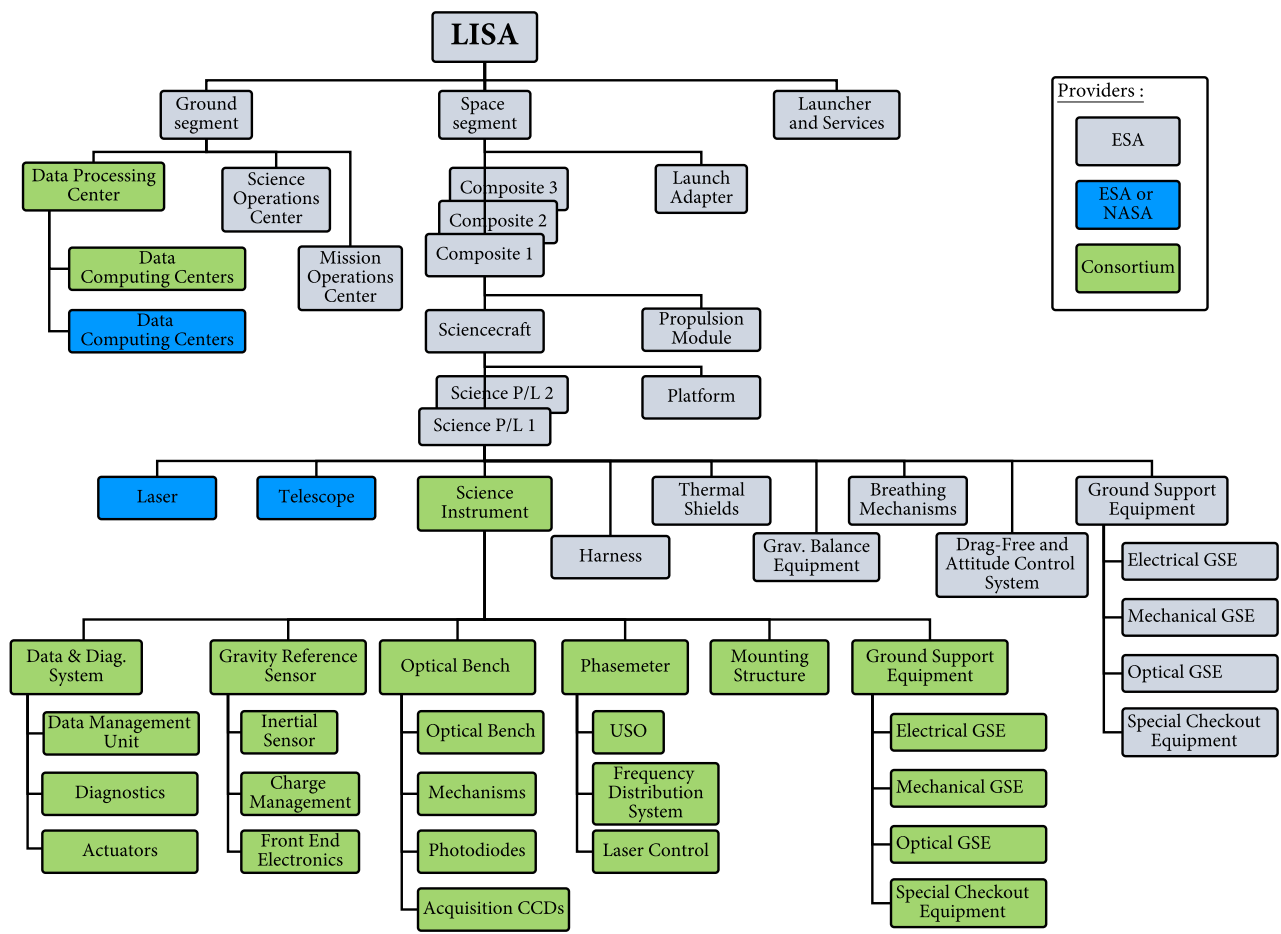

**Figure 14: A 'strawman' representation of elements and assignments of responsibilities**, to be refined during Phase-A.

## 8.2 Payload and Instrument Description

**Product Tree**

A 'strawman' product tree, together with assigned responsibilities, is shown in Figure 14.

*Space Segment* The space segment covers all the activities that are necessary to build, test and deliver a fully functional launch composite that is compliant to the requirements. The composite comprises the sciencecraft (made of S/C bus and P/L) and the propulsion module. This element also includes the Ground Support Equipments (GSE) and the Real-time Test Bed (RTB) required for ground testing. The ESA mission prime is responsible for the composite, based on the provision of the Instrument by the Member State Consortium. Each S/C comprises the S/C bus and the payload.

From the organisational point of view, one prime con-tractor, three major subcontractors (AIVT, propulsion, P/L), subsystem and equipment contractors for the S/C bus elements are foreseen. The ESA payload contractor procures the laser and the telescope to the payload, or receives these items from NASA. The contractor is responsible for the specification, procurement, assembly, integration, verification, and testing of the complete payload. The LISA Consortium will deliver the Instrument to the payload contractor.

*Launcher and Services* This element includes the Ariane 6.4 launcher and the services at the launch site. All three LISA S/C are planned to be launched together by an Ariane 6.4 launch vehicle. The Ariane launch will deliver the composites into a direct escape hyperbola, from which they will individually reach their final orbital position by means of a chemical propulsion module that will be jettisoned before starting the scientific operations.

*Ground Segment* The Ground Segment element includes all Mission Operations during Low Earth Operations (LEOP), transfer and later during nominal operations and Science Operations under ESA responsibility, that is raw science data pre-processing and calibration, leading to level-1 data (TDI combinations). This task will be performed with support from France, Italy, the United Kingdom, Switzerland, Spain, Germany and the US (algorithm development) and the instrument providers (calibration during operation).

### 8.2.1 Model Philosophy

Following the LISA Pathfinder strategy, the proposed model philosophy of the Science Instruments is based on the development of one Structural and Thermal Model (STM), one Engineering Model (EM), one Proto-Flight Model (PFM) and 5 additional Flight Models. This development philosophy is expected to be studied and re-assessed during Phase A.

Optionally, a spare unit of the instrument might also be developed. A preferred option is to keep provision of spare parts, components and boards in each supplier, so that any failed electronic devices may be repaired and tested within a few weeks (similar to the option retained for LISA Pathfinder). During the AIVT flow, the engineering phase consists in the development of 2 models: the STM (Structural and Thermal Model) and the Engineering model (EM) of the Science Instrument. One proto-flight (PFM) and five flight (FM) models are developed afterwards. The qualification tests are performed on the PFM, while the FMs undergo acceptance tests. As a general rule, the PFM and FM AIT flows follow the one defined for the EM, in order to benefit from the tests procedure, performance results and developed GSE and Special Check-Out Equipments (SCOE).

### 8.2.2 Deliverables to the instrument integrator

The Consortium members are responsible for delivering the agreed models (STM, EM, FM etc.) of their own equipment (H/W and S/W as needed) to the AIVT manager in the Instrument Management Team according to the agreed development schedule. The individual command/control electronics, S/W, user's manual, performance checks, mechanical drawings, etc. shall also be delivered in order to allow acceptance tests and later performance tests of the equipment by the AIVT manager.

In addition to the agreed equipment models, the Consortium members will also deliver the associated SCOEs and simulators needed for the instrument integration.

The AIVT manager is responsible for developing and maintaining the appropriate ground support equipment (H/W and S/W) needed to integrate, test and validate the system performance of the instrument.

## 8.3 International Partners

LISA has a long history of joint development between European and US scientists. In fact, the original LISA proposal was assuming an equal partnership between ESA and NASA on LISA. In 2011 that plan had to be abandoned and the L1-NGO proposal was based on a pure ESA mission with European member state participation. This has now been superseded by a considerable evolution of the NASA position and a strong endorsement by the mid-Decadal Review in the US. This proposal is based on the assumption that NASA will participate in an ESA-led LISA mission at the level of about 20% of the total mission cost. The specific items and mission elements to be provided by NASA will of course be subject to agency-level negotiations. But preliminary discussion at the scientific working level have helped to identify potentially promising items that could greatly increase the science output of the mission.

The specific items identified so far comprise the following potential contributions to the LISA payload directly supplied by NASA to ESA:

- Space-qualified laser systems
- Frequency reference cavity for laser stabilisation
- Send/receive telescopes

Potential contributions to the S/C that could be made by NASA to ESA:

- Propulsion modules
- Solar panels
- Micropropulsion systems

NASA may also contribute elements of the LISA instrument to the European member states, such as:

- Charge management system
- Optical bench photoreceivers and front-end electronics
- Contributions to phasemeter hardware and software

### 8.4 Preliminary Program Schedule

The LISA schedule can be fully compatible with the L3 schedule quoted in the call for mission concepts. But after the mission success of LISA Pathfinder we are convinced that an accelerated schedule for LISA is technically possible with no additional risk. Many technology items are now flight ready and need little additional development. The remaining technology developments are expected to be finished by 2020 such that all technologies are at TRL 6 or higher. Following purely technical readiness, a launch before 2030 is feasible, as is shown by this schedule produced by the GOAT [47] committee (see Figure 15) as an example of what would be possible without financial or programmatic constraints.

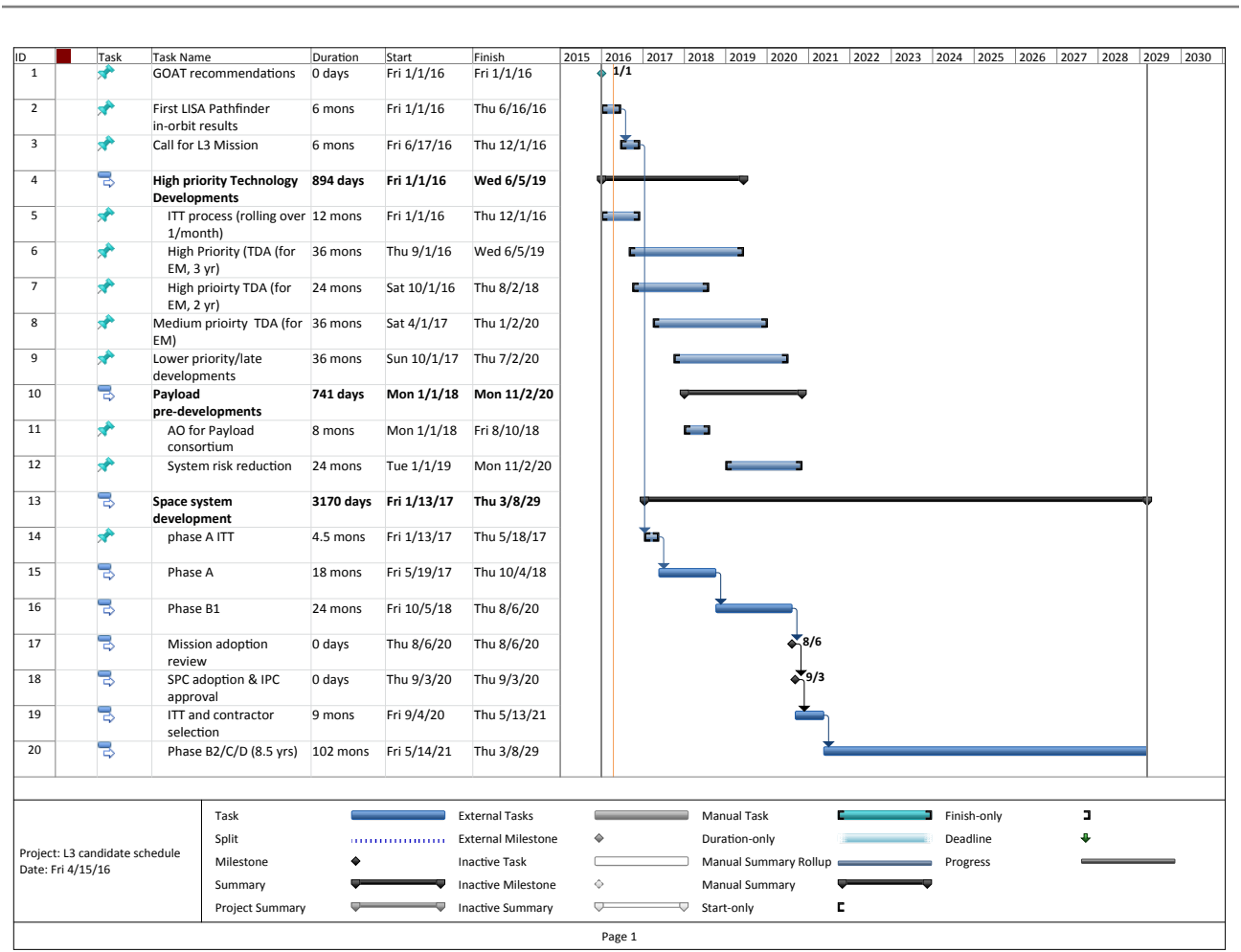

Figure 15: The schedule developed as part of the GOAT committee exercise.

### 8.5 Preliminary Cost Analysis

The detailed analysis of mission cost will be the subject of the upcoming studies. Nevertheless, building on the considerable heritage of more than 10 years of Mission Formulation study for LISA and a cost assessment by ESA and a grass-roots costing by the member state agencies for the L1-NGO proposal, we can give a reasonably reliable estimate that makes it plausible that our proposed mission concept for LISA will fulfill all science requirements and can be flown within the cost envelope of an L3 mission.

The total payload cost for L1-NGO was assessed as 175

M€ in 2012 economic conditions. Since our baseline mission now entails 6 laser links instead of the 4 assumed for NGO, the recurring cost for some member states will be somewhat higher, and we estimate the total cost to the member states at 250 M€ without margin. This seems within the affordable range for the ESA member states.

The ESA cost at completion (CaC) for the NGO layout as a strawman design for the L2/L3 Science Theme Selection was costed by ESA in 2013 as 1200 M€, assuming an Ariane 6 launch and procurement of lasers and telescopes by ESA. Our current LISA proposal is built on using three identical S/C of the type of the NGO mother S/C, so we can rely on scaling the mass, volume and cost from those numbers. The main differences come from the slightly larger telescopes and the slightly changed orbits. Increasing the telescope diameter from 20 cm to 30 cm should not change the S/C mass and cost significantly, because that is mainly determined by the optical bench size as long as the telescope diameter stays below 30 cm. The NGO orbits were drift-away orbits, limiting the possible extended mission lifetime. We are now asking for a stop manoeuvre at the end of the orbit insertion to lower Doppler shifts and breathing angles, reducing payload complexity and risk, and making a 10 year extended mission feasible. We note that this only adds a $\Delta V$ penalty of a few 100 m/s and needs very little additional fuel, which is well within the Ariane 6.4 capabilities. So we expect no noticeable cost penalty from this. Moreover, since we are asking for three identical S/C for our baseline mission, the additional non-recurring cost of having two different S/C types in NGO no longer applies to our LISA design. And finally, if NASA provides, for example, lasers, telescopes and some other spacecraft and instrument elements, we expect that the cost to ESA will stay well below the 1050 M€ cost cap for L3 missions.

## 8.6 Risk analysis

Risk items were identified in the technical and programmatic review of the NGO L1 Mission Study [48], page 12-14. After the successful pathfinder flight, many of the risk items can be retired.

The top-level (D4 according to ECSS-M-ST-80C [49]) risk (Complexity caused by multiple spacecraft development will impact schedule and cost), due to system complexity, can now be reduced to a C4 because a) we are using 3 identical spacecraft instead of two types and b) due to the successful integration of LPF and the demonstration of gravity gradient control. The next risk (C4) (Inability to perform end-to-end testing on the ground will result in degraded mission capabilities) can be reduced to a B4 due to the end-to-end demonstration of success in LPF. The 3$^{rd}$ item (B5)(Failure of a single GRS system degrades science performance) is retired to B3 because we have 3 instead of 2 arms, and the GRS has direct flight heritage. There is no quantifiable change to the risk that loss of one S/C will cause the end of mission. But we intend to study the option to assemble a full spacecraft flight spare on the ground from the various subsystem flight spares. In case of a spacecraft failure during the extended mission this could be launched on demand.

The risk that LPF will fail to demonstrate some in-flight performance at the required levels or the data cannot be extrapolated to LISA performance is retired. The (formerly C3) Cold Gas Thrusters have been qualified for the 5-year Gaia performance lifetime. The failure of LPF has been completely retired. The risk (B5) "Acquisition of the optical links through the telescopes between spacecraft not achieved" is reduced through the performance of the differential wavefront sensing on LPF to B3.

The risk "PM Separation results in high spacecraft rotation rate" has been retired by test on LPF. The risk that there will be "Unexpected thermal fluctuation noise sources degrade residual acceleration performance" is greatly reduced through LPF performance to A2. The risk that "Transfer Burns fail to insert spacecraft into final orbits" is partly retired due to a successful test on LPF.

Remaining notable risks for LISA are under control due to the ongoing LISA development program. They are: the risk that two different architectures are possible for the telescope pointing; optical bench series production; straylight and manufacturability of the telescope; and laser system sideband fidelity.

Lastly, we note the risk that the long program duration leads to loss of key and experienced personnel. This is best mitigated through an early implementation and early launch date, and continued robust support from the contributing agencies. This will permit transfer of knowledge in the scientific instrument teams from one generation of scientists to the next during the mission lifetime. Assuming early and reliable support from our national agencies, we are planning to develop a personnel succession plan for younger scientists supporting the key people close to retirement.

# 9 Conclusion

The groundbreaking discovery of Gravitational Waves by ground-based laser interferometric detectors in 2015 has changed astronomy, by giving us access to the high-frequency regime of Gravitational Wave astronomy. By 2030 our understanding of the Universe will have been dramatically improved by new observations of cosmic sources through the detection of electromagnetic radiation and high-frequency Gravitational Waves. But in the low-frequency Gravitational Wave window, below one Hertz, we expect to observe the heaviest and most distant objects. Using our new sense to 'hear' the Universe with LISA, we will complement our astrophysical knowledge, providing access to a part of the Universe that will forever remain invisible with light. LISA will be the first ever mission to survey the entire Universe with Gravitational Waves. It will allow us to investigate the formation of binary systems in the Milky Way, detect the guaranteed signals from the verification binaries, study the history of the Universe out to redshifts beyond 20, when the Universe was less than 200 million years old, test gravity in the dynamical sector and strong-field regime with unprecedented precision, and probe the early Universe at TeV energy scales. LISA will play a unique and prominent role in the scientific landscape of the 2030s.

Front Cover Image Credits:

- Background: a composition of the center of the milky way (custom composition of three different wavelengths images) and a deep star map by NASA's scientific visualization studio
- Earth: textures are from NASA blue marble, 3D rendering from Simon Barke
- LISA constellation: Simon Barke